\documentclass[11pt, twocolumn]{IEEEtran}

\usepackage[utf8]{inputenc}
\usepackage[T1]{fontenc}

\usepackage[cmex10]{amsmath}
\usepackage{amssymb}
\usepackage{amsthm}
\usepackage{bm}

\usepackage{dsfont}

\usepackage{mleftright}

\usepackage{bbm} 

\mleftright

\usepackage{algorithm}
\usepackage{algpseudocode}

\usepackage{graphicx}
\usepackage[space]{grffile}
\usepackage{booktabs}

\usepackage{tikz}
\usetikzlibrary{arrows.meta,automata,positioning,fit}

\usepackage{quantikz}

\usepackage{cite}
\usepackage{url}
\usepackage{enumitem}
\usepackage{xcolor}

\theoremstyle{definition}

\newtheorem{definitionenv}{Definition}
\newtheorem{lemmaenv}[definitionenv]{Lemma}
\newtheorem{theoremenv}[definitionenv]{Theorem}
\newtheorem{corollaryenv}[definitionenv]{Corollary}
\newtheorem{propositionenv}[definitionenv]{Proposition}
\newtheorem{conjectureenv}[definitionenv]{Conjecture}
\newtheorem{remarkenv}[definitionenv]{Remark}
\newtheorem{exampleenv}[definitionenv]{Example}

\newenvironment{definition}{\begin{definitionenv}\rm}{\end{definitionenv}}
\newenvironment{lemma}{\begin{lemmaenv}\rm}{\end{lemmaenv}}
\newenvironment{theorem}{\begin{theoremenv}\rm}{\end{theoremenv}}
\newenvironment{corollary}{\begin{corollaryenv}\rm}{\end{corollaryenv}}
\newenvironment{proposition}{\begin{propositionenv}\rm}{\end{propositionenv}}

	\newcommand{\ed}{\end{definition}}
	\newcommand{\el}{\end{lemma}}
\newcommand{\bt}{\begin{theorem}}
	\newcommand{\et}{\end{theorem}}
\newcommand{\bc}{\begin{corollary}}
	\newcommand{\ec}{\end{corollary}}
	\newcommand{\ep}{\end{proposition}}

\newcommand{\cL}{{\cal L}}

\newcommand{\cQ}{{\cal Q}}
\newcommand{\cS}{{\cal S}}
\newcommand{\cT}{{\cal T}}
\newcommand{\cP}{\mathcal{P}}
\newcommand{\cH}{\mathcal{H}}
\newcommand{\cV}{\mathcal{V}}
\newcommand{\cF}{\mathcal{F}}
\newcommand{\cX}{\mathcal{X}}

\newcommand{\mba}{{\mathbf{a}}}
\newcommand{\mbb}{{\mathbf{b}}}
\newcommand{\mbc}{{\mathbf{c}}}

\newcommand{\mbz}{{\mathbf{z}}}

\newcommand{\mbh}{{\mathbf{h}}}

\newcommand{\mbr}{{\mathbf{r}}}
\newcommand{\mbs}{{\mathbf{s}}}
\newcommand{\mbt}{{\mathbf{t}}}
\newcommand{\mbu}{{\mathbf{u}}}
\newcommand{\mbv}{{\mathbf{v}}}
\newcommand{\mbw}{{\mathbf{w}}}
\newcommand{\mbx}{{\mathbf{x}}}
\newcommand{\mby}{{\mathbf{y}}}

\newcommand{\wt}{\mathrm{wt}}
\newcommand{\Tr}{\mathrm{tr}}

\usepackage[colorinlistoftodos,prependcaption]{todonotes}

\usepackage{xargs}

\newcommandx{\rednote}[2][1=]{\todo[inline,linecolor=red,backgroundcolor=red!25,bordercolor=red,#1]{#2}}
\newcommandx{\yellownote}[2][1=]{\todo[inline,linecolor=yellow,backgroundcolor=yellow!25,bordercolor=yellow,#1]{#2}}

\begin{document}

\title{Twisted Fourier Analysis of Quantum Weight Enumerators}

\author{Ching-Yi~Lai
\thanks{\footnotesize
CYL was supported by the National Science and Technology Council in Taiwan under Grant Nos. 113-2221-E-A49-114-MY3 and 113-2119-M-A49-008-.

CYL is with the Institute of Communications Engineering,
National Yang Ming Chiao Tung University,
Hsinchu 300093, Taiwan
(email: cylai@nycu.edu.tw).}}

\maketitle

\begin{abstract}
	
	A unified Fourier framework for
	quantum weight enumerators of prime-power-dimensional qudits
	is developed in this paper.
	We introduce the notions of Weyl power
	spectrum and Weyl-conjugation autocorrelation
	for the Weyl-operator quantities underlying
	the Shor--Laflamme weight enumerators.
	Representing these quantities as formal vectors,
	we show that they form a symplectic Fourier
	transform pair, from which the quantum
	MacWilliams identity follows by applying
	the Hamming-weight functional.
	This framework extends to a family
	of twisted Fourier
	transforms, including the Rains shadow transform.
	We show that identical-copy Bell sampling
	corresponds to a twisted autocorrelation
	related to the shadow enumerator through
	Weyl translation and Hamming projection.
	We further prove that all twisted Fourier transforms
	are Weyl translations of either the ordinary
	MacWilliams transform or the Rains shadow transform.
	This two-class reduction implies that
	the twists yield no additional independent
	coefficientwise positivity constraints
	under compatible projections.
	Nevertheless, the framework leads to
	refined MacWilliams and shadow identities
	that retain more Fourier information than
	ordinary Hamming weight enumerators.
	We employ these refined identities in
	linear programming, together with
	simplex mapping and shadow-coset counting,
	to exclude surviving candidate weight
	distributions.
	These techniques yield tighter bounds
	for binary and qutrit quantum codes,
	including nonexistence results for
	specific code parameters.

\end{abstract}

\section{Introduction}
\label{sec:introduction}

Determining the fundamental limits of
quantum error-correcting codes is a
central problem in quantum coding theory~\cite{CRSS98}.
Shor and Laflamme~\cite{SL97} introduced
the $A$- and $B$-type weight enumerators
based on Pauli operators and derived
a quantum analogue of the classical
MacWilliams identity~\cite{MS77}
through the relation between entanglement
fidelity and average fidelity
under depolarizing noise.
This identity enables the extension of
Delsarte's classical linear programming
(LP) method~\cite{Del73} to quantum
code bounds~\cite{AL99}.
Rains subsequently developed the general theory
of quantum weight enumerators~\cite{Rains98a}
and introduced the shadow enumerator~\cite{Rains99b},
which provides additional LP constraints.
Further interpretations of quantum weight
enumerators have been investigated
in~\cite{NK22,LTY24}.
Weight-enumerator methods have also been
extended to subsystem codes~\cite{AKS06},
entanglement-assisted quantum codes~\cite{LBW13,LBW14},
codes for simultaneous transmission of
classical and quantum information~\cite{GLZ17},
quantum data-syndrome codes~\cite{ALB20},
quantum codes under general noise channels~\cite{OL22},
tensor networks~\cite{cao2023quantum},
and magic state distillation~\cite{rall2017signed,kalra2025invariant}.
Beyond weight-enumerator LP methods,
semidefinite programming (SDP) techniques
have also been developed to obtain
stronger bounds on quantum code parameters
~\cite{LTY24,MNH24}.

In this paper, we develop a unified Fourier
formulation of quantum weight enumerators
for prime-power-dimensional qudit systems.
We introduce the notions of Weyl power spectrum
and Weyl-conjugation autocorrelation
for the Weyl-operator quantities underlying
the Shor--Laflamme weight enumerators,
in analogy with the classical
Wiener--Khinchin theorem~\cite{Khi34}.
By representing these quantities as formal
vectors indexed by elements of $\mathbb F_q^{2n}$,
we show that they form a symplectic Fourier
transform pair~\cite{LHL16}.
Applying the Hamming-weight functional,
which assigns $x$ and $y$ to zero and
nonzero Weyl labels, respectively,
recovers the conventional quantum
MacWilliams identity~\cite{SL97}.

This framework can be extended to
$2q^2$ twisted Fourier transforms
by introducing different character weights
on Weyl labels.
For qubits, these comprise eight transforms,
including the ordinary MacWilliams~\cite{SL97}
and Rains shadow~\cite{Rains99b} transforms,
together with six additional Pauli twists.

These twisted transforms also admit
operational interpretations through Bell sampling.
Recent works have established connections
between quantum weight enumerators,
shadow enumerators, and two-copy
Bell measurements~\cite{SGZZ26,MLP+26}.
We clarify this connection by showing that
identical-copy Bell sampling corresponds
to the pointwise $010$ twisted autocorrelation,
rather than directly to the shadow transform.
A Weyl translation followed by Hamming
projection then recovers the Rains shadow
enumerator, as illustrated in
Figure~\ref{fig:bell_shadow_relation}.

More generally, we show that for
$q=p^m$, all $2q^2$ twisted Fourier
transforms are Weyl translations of either
the ordinary MacWilliams transform or
the Rains shadow transform.
This two-class reduction implies that
these twists yield no additional independent
coefficientwise positivity constraints
at the full Weyl-resolved level.

A natural question is which projections
preserve the symplectic Fourier transform.
We introduce the qudit refinement $R_\eta$,
which partitions the $q^2$ Weyl labels
into $2q-1$ classes, and prove its
invariance under the symplectic Fourier
transform.
For qubits, this reduces to the $R_X$
refinement, which distinguishes $I$, $X$,
and $\{Y,Z\}$~\cite{CRSS98,LA18}.
We also develop its split form
$H\times R_\eta$, which retains more
Fourier information than ordinary Hamming
weight enumerators while requiring
substantially fewer variables than the
complete Weyl enumerator.
For $[[n,k,d]]_q$ stabilizer codes~\cite{Got97,KL97,KKKS06},
fixing a reference stabilizer further
introduces finite-field commutation constraints,
multiplication orbits, and divisibility
conditions.
Together, these properties yield refined
MacWilliams and shadow identities and
additional LP constraints that can improve
ordinary Hamming-level bounds.
As an illustration, we use the refined LP
constraints to rule out quantum codes with
parameters $[[11,3,5]]_3$ and $[[11,5,4]]_3$,
thereby tightening the upper bounds listed
in Grassl's code tables~\cite{Grassl}.
These nonexistence results have also been
obtained recently through a classification-based
approach~\cite{BGL23}.

Although the refined LP constraints can be
substantially stronger than ordinary
Hamming-level bounds, they are computationally
more demanding.
In practice, we first use the
Shor--Laflamme--Rains (SLR) LP constraints
and the three-case counting
conditions of~\cite[Theorem 16]{LA18},
referred to as SLR3,
to obtain a smaller set of candidate
weight distributions before applying
the refined LP.
Since LP-feasible weight distributions
need not be realizable, we develop
additional combinatorial constraints.

For $[[17,6,5]]$, the SLR3 enumeration
leaves 292 candidate weight distributions.
We develop shadow-coset counting constraints
based on the low-weight structure of
Pauli shadow cosets and show that they
exclude all 292 candidates, thereby
establishing the nonexistence of such a code.

We also employ the simplex mapping~\cite{CRSS98}
to obtain further nonexistence results.
Classical linear complementary dual (LCD)
codes have been studied extensively
~\cite{Massey92,DKO+17,Bouyuklieva21,LuLiRen24}
and are closely related to
$[[n,k,d;n-k]]$ entanglement-assisted (EA)
stabilizer codes with maximal entanglement
~\cite{LBW13,LBW14,LuLiGuoFu15}.
In the reverse direction, we show that
an $[[n,k,d;n-k]]$ EA code yields a
$[3n,2k,2d]$ binary LCD code through
the simplex mapping.
Applying classical LCD bounds, we rule out
four maximally entangled binary EA code
parameter sets:
$[[12,6,6;6]]$, $[[15,4,10;11]]$,
$[[17,13,4;4]]$, and $[[19,3,14;16]]$.

Finally, for $[[14,3,5]]$ stabilizer codes,
the simplex mapping converts the stabilizer group
into a $[42,11]$ binary linear code.
By analyzing the low-weight structure
of its binary shadow coset,
we exclude the remaining candidate
weight distributions and establish
the nonexistence of such codes.

The nonexistence results obtained through
refined LP, simplex mapping, and
shadow-coset counting are summarized
in Table~\ref{tab:nonexistence}.

\begin{table}[t]
	\centering
	\caption{Nonexistence results obtained by the
		methods developed in this paper.}
	\label{tab:nonexistence}
	\begin{tabular}{ll}
		\toprule
		Code parameters & Method \\
		\midrule
		$[[11,3,5]]_3$ & Refined LP \\
		$[[11,5,4]]_3$ & Refined LP \\
		\midrule
		$[[17,6,5]]$ & SLR3 + Pauli shadow-coset counting \\
		$[[14,3,5]]$ & Simplex mapping + binary shadow-coset counting \\
		\midrule
		$[[12,6,6;6]]$ & Simplex mapping + LCD bounds \\
		$[[15,4,10;11]]$ & Simplex mapping + LCD bounds \\
		$[[17,13,4;4]]$ & Simplex mapping + LCD bounds \\
		$[[19,3,14;16]]$ & Simplex mapping + LCD bounds \\
		\bottomrule
	\end{tabular}
\end{table}

The source code and candidate weight
distributions used in our computational
analysis are publicly available
on GitHub~\cite{Lai2026QuantumCodeLP}.

The remainder of this paper is organized
as follows.
Preliminaries on Weyl operators over finite fields and the symplectic Fourier transform
are given in the next section.
In Section~\ref{sec:qudit_pps}, we introduce
the Weyl power spectrum and
Weyl-conjugation autocorrelation, and show
that they form a symplectic Fourier
transform pair.
We then develop twisted Fourier transforms
in Section~\ref{sec:qudit_twists} and prove
their two-class reduction in
Section~\ref{sec:twist_redundancy}.
The refined LP constraints and computational
results are presented in Section~\ref{sec:LP}.
In Section~\ref{sec:structural}, we further
develop simplex representations and
shadow-coset counting methods to exclude
additional candidate weight distributions.
Finally, the conclusion and discussion
of open problems follow.

\section{Qudit Weyl Phase Space}
\label{sec:qudit_prelim}

For a set \(C\), let \(\mathds{1}_{C}\) denote its indicator
function, defined by
\begin{equation*}
	\mathds{1}_{C}(x)
	=
	\begin{cases}
		1,& x\in C;\\
		0,& x\notin C.
	\end{cases}
\end{equation*}
For $x,y$ in the same domain, denote by
$\mathds{1}_{\{x=y\}}$ the indicator of the equality $x=y$.

Let $[n]=\{1,2,\ldots,n\}$.
For subsets $T,U\subseteq[n]$,
the notation $T\sqcup U=[n]$ means
$T\cup U=[n]$ and $T\cap U=\emptyset$.
When $|T|=t$ and $|U|=u$,
we may assume without loss of generality that
$T=\{1,\ldots,t\}$ and
$U=\{t+1,\ldots,n\}$.

\subsection{Finite-field characters and the trace-symplectic form}

We first review some standard notation for Weyl operators and their associated phase spaces and additive characters (see, e.g., \cite{KKKS06,terras1999fourier}).

Let $q=p^m$ be a prime power, where $p$ is prime. Denote by $\mathbb F_q$ the finite field with $q$ elements and by $\mathbb F_q^\times$ its multiplicative group of nonzero elements.
The field trace
\(\operatorname{Tr}_{\mathbb F_q/\mathbb F_p}:\mathbb F_q\to\mathbb F_p\)
is defined by
$
	\operatorname{Tr}_{\mathbb F_q/\mathbb F_p}(x)	=
	\sum_{i=0}^{m-1}x^{p^i}.
$
Let \(\omega=e^{2\pi i/p}\) be a primitive \(p\)-th root of unity. The canonical additive character of
\(\mathbb F_q\) is
\begin{equation}
	\chi_q(x)
	=
	\omega^{\operatorname{Tr}_{\mathbb F_q/\mathbb F_p}(x)}.
\end{equation}

The single-qudit Weyl phase space is
\(\mathcal V_q=\mathbb F_q^2\).
For \(u=(a,b)\) and \(v=(a',b')\) in \(\mathcal V_q\), define the
trace-symplectic form
\begin{equation}
	\langle u,v\rangle
	=
	\operatorname{Tr}_{\mathbb F_q/\mathbb F_p}(ba'-ab')
	\in\mathbb F_p,
\end{equation}
and for \(\mbu=(u_1,\ldots,u_n),\mbv=(v_1,\ldots,v_n)\in\mathcal V_q^n\), set
\(\langle\mbu,\mbv\rangle=\sum_{j=1}^n\langle u_j,v_j\rangle\).

An additive character of \(\mathcal V_q^n\) is a homomorphism from its
additive group to the unit circle in \(\mathbb C\).
For each \(\mbu\in\mathcal V_q^n\), the associated character is
\begin{equation}
	\chi_\mbu(\mbv)
	=
	\omega^{\langle\mbu,\mbv\rangle}.
\end{equation}
These characters satisfy the standard orthogonality relation
\begin{equation}
	\sum_{\mbv\in\mathcal V_q^n}\chi_\mbu(\mbv)
	=\sum_{\mbv\in\mathcal V_q^n}\omega^{\langle\mbu,\mbv\rangle}=
	q^{2n}\,\mathds{1}_{\{\mbu=\mathbf{0}\}}.  \label{eq:character_orthogonality}
\end{equation}

\subsection{Weyl operators}

Let \(\cH_q=\mathbb C^q\) denote the Hilbert space of a \(q\)-dimensional qudit, with computational basis \(\{\ket{x}:x\in\mathbb F_q\}\). The \(n\)-qudit state space is \(\cH_q^{\otimes n}\). Let \(\cL(\cH_q^{\otimes n})\) denote the space of linear operators on \(\cH_q^{\otimes n}\), equipped with the Hilbert--Schmidt inner product
\(\langle O_1,O_2\rangle_{\mathrm{HS}}=\Tr(O_1^\dagger O_2)\) for \(O_1,O_2\in\cL(\cH_q^{\otimes n}),\)
where \(\Tr\) denotes the usual operator trace.

For \(a,b\in\mathbb F_q\), define
\begin{align*}
	X(a)|x\rangle&=|x+a\rangle,\\
	Z(b)|x\rangle&=\chi_q(bx)|x\rangle.
\end{align*}

We use the phase convention \(W_u=W_{a,b}=X(a)Z(b)\).
The operators satisfy
\(Z(b)X(a)=\chi_q(ab)X(a)Z(b)\),
and hence, for \(u,v\in\mathcal V_q\),
\begin{equation}W_uW_vW_u^\dagger=\omega^{\langle u,v\rangle}W_v.
\end{equation}

For \(\mbu=(u_1,\ldots,u_n)\in\mathcal V_q^n\), define the \(n\)-qudit Weyl operator by
\(W_\mbu=\bigotimes_{j=1}^n W_{u_j}\).
The conjugation relation then extends to \(\mathcal V_q^n\): for \(\mbu,\mbv\in\mathcal V_q^n\),
\begin{equation}
	W_\mbu W_\mbv W_\mbu^\dagger
	=
	\omega^{\langle\mbu,\mbv\rangle}W_\mbv. \label{eq:Weyl_conjugation}
\end{equation}

The Weyl operators form an orthogonal operator basis  for \(\cL(\cH_q^{\otimes n})\) :
\begin{align}
	\Tr(W_\mbu^\dagger W_\mbv)
	=
	q^n\mathds{1}_{\{\mbu=\mbv\}}.
	\label{eq:weyl_orthogonality}
\end{align}

The (symplectic Hamming) weight of \(\mbv=(v_1,\ldots,v_n)\in\mathcal V_q^n\) is
$
	\wt(\mbv)=|\{j:v_j\neq(0,0)\}|.
$
That is, $\wt(\mbv)$ is the number of nonidentity tensor factors
in $W_\mbv$. We also define $\wt(W_\mbv)=\wt(\mbv)$.

For $q=2$, $\mathcal V_2=\mathbb F_2^2$ has four
elements $00,10,01,11$, corresponding to the
Pauli operators $I,X,Z,Y$, respectively, up to phases.
The symplectic form
$\langle\mbu,\mbv\rangle$ equals $0$ if the
corresponding Pauli operators commute and $1$
if they anticommute.
In this case, $\omega=-1$, and the conjugation
phase reduces to $(-1)^{\langle\mbu,\mbv\rangle}$,
the usual Pauli commutation sign.


\subsection{The Fourier operator}

We introduce a Hilbert-space representation of the Weyl phase space and the associated Fourier transform, following~\cite{LHL16}. This representation provides the vector notation used throughout the paper.

Let \(\cH_{\cV_q}\) be the formal Hilbert space with orthonormal basis \(\{\ket{v}:v\in\cV_q\}\). The unnormalized single-qudit  Fourier operator is defined as
\begin{align}
	\cF_q=\sum_{u,v\in\cV_q}
	\omega^{\langle u,v\rangle}\ket{u}\bra{v}.
	\label{eq:qudit_fourier_operator}
\end{align}
It is clear that \(\langle\mbv,\mbu\rangle= - \langle\mbu,\mbv\rangle\) and thus  $ \cF_q^\dagger=\cF_q$.
By~\eqref{eq:character_orthogonality}, \(\cF_q^2=q^2 I\). Thus \(q^{-1}\cF_q\) is a Hermitian unitary operator.

For \(n\) qudits, \(\cH_{\cV_q^n}=(\cH_{\cV_q})^{\otimes n}\), and the corresponding symplectic Fourier operator is \(\cF_q^{\otimes n}\).

\subsection{Quantum codes and the QEC condition}

By~\eqref{eq:weyl_orthogonality}, every operator \(O\in\cL(\cH_q^{\otimes n})\) admits the Weyl expansion
\begin{align}
	O
	=
	\frac{1}{q^n}
	\sum_{\mbu\in\cV_q^n}
	\Tr(W_\mbu^\dagger O)\,W_\mbu.
	\label{eq:operator_weyl_expansion}
\end{align}

An \(((n,M,d))_q\) quantum code \(\cQ\) is an \(M\)-dimensional subspace of \(\cH_q^{\otimes n}\). Let \(P\) denote the projector onto \(\cQ\), so that \(\Tr(P)=\operatorname{rank}(P)=M\). The minimum distance \(d\) is the largest integer such that every Weyl error of weight less than \(d\) is detectable~\cite{KL97,KKKS06}; that is, for every \(\mbu\in\cV_q^n\) with \(\wt(\mbu)\le d-1\),
\begin{align}
	PW_\mbu P
	=
	c_\mbu P,
	\label{eq:QEC_condition}
\end{align}
where \(c_\mbu\in\mathbb C\).
Equivalently, for any code states \(\ket{\psi},\ket{\phi}\in\cQ\),
\begin{align*}
	\langle\phi|W_\mbu|\psi\rangle
	=
	c_\mbu\langle\phi|\psi\rangle.
\end{align*}
In particular, the error does not confuse
distinct orthogonal code states.
When $M=q^k$, the code is denoted by
$[[n,k,d]]_q$, with the subscript omitted
for $q=2$.

\section{Weyl Power Spectrum and the Qudit MacWilliams Identity}
\label{sec:qudit_pps}

Let \(\cQ\subset\cH_q^{\otimes n}\) be an \(((n,M,d))_q\) quantum
code with projector \(P\), so that \(\Tr(P)=M\).
By the Weyl expansion~\eqref{eq:operator_weyl_expansion},
\begin{align}
	P=\frac{M}{q^n}\sum_{\mbv\in \cV_q^n}\lambda(\mbv)W_\mbv,
	\label{eq:qudit_projector_expansion}
\end{align}
where
\begin{align}
	\lambda(\mbv)
	=\frac{1}{M}\Tr(W_\mbv^\dagger P),
	\label{eq:qudit_lambda}
\end{align}
is the normalized Weyl coefficient. These coefficients describe the components of the code projector in the Weyl operator basis. We next associate with them two quantities: the \textit{Weyl power spectrum}, which records the squared magnitudes of the Weyl coefficients, and the \textit{Weyl-conjugation autocorrelation}, which measures the overlap of the code projector with its Weyl conjugates.

\begin{definition}
	The Weyl power spectrum of  \(P\) is defined as
	\begin{align}
		\Lambda(\mbv)=|\lambda(\mbv)|^2
		=\frac{1}{M^2}
		\Tr(W_\mbv^\dagger P)\Tr(W_\mbv P),
		\label{eq:qudit_power_spectrum}
	\end{align}
	and its formal vector is
	\begin{align}
		|\Lambda\rangle
		=\sum_{\mbv\in \cV_q^n}\Lambda(\mbv)|\mbv\rangle.
	\end{align}
\end{definition}

\begin{definition}
	The Weyl-conjugation autocorrelation of \(P\) is defined as
	\begin{align}
		\widehat\Lambda(\mbu)
		=\frac{1}{M}\Tr(W_\mbu P W_\mbu^\dagger P),
		\label{eq:qudit_dual_spectrum}
	\end{align}
	and its formal vector is
	\begin{align}
		|\widehat\Lambda\rangle
		=\sum_{\mbu\in \cV_q^n}\widehat\Lambda(\mbu)|\mbu\rangle.
	\end{align}
\end{definition}

The two quantities are not independent. As shown below, the Weyl-conjugation autocorrelation is precisely the  Fourier transform of the Weyl power spectrum.

\begin{theorem}[Weyl power-spectrum MacWilliams identity]
	\label{thm:qudit_pps_MI}
	For an \(((n,M))_q\) quantum code \(\cQ\) with projector \(P\), let \(\Lambda\) and \(\widehat\Lambda\) be its Weyl power spectrum and Weyl-conjugation autocorrelation, respectively. Then
	\begin{align}
		\ket{\widehat\Lambda}
		=
		\frac{M}{q^n}\cF_q^{\otimes n}\ket{\Lambda}.
		\label{eq:qudit_pps_MI}
	\end{align}
	Equivalently,
	\begin{align}
		\widehat\Lambda(\mbu)
		=
		\frac{M}{q^n}
		\sum_{\mbv\in\cV_q^n}
		\omega^{\langle\mbu,\mbv\rangle}\Lambda(\mbv).
		\label{eq:qudit_pps_MI_coeff}
	\end{align}
	Moreover, for every \(\mbv\in\cV_q^n\),
	\begin{align}
		\widehat\Lambda(\mbv)\ge \Lambda(\mbv)\ge0.
		\label{eq:qudit_pointwise_AB}
	\end{align}
\end{theorem}

\begin{proof}
	Taking the Weyl conjugation~\eqref{eq:Weyl_conjugation} of \(P\) by \(W_\mbu\) in the expansion~\eqref{eq:qudit_projector_expansion}, we have
	\begin{align*}
		W_\mbu P W_\mbu^\dagger
		=
		\frac{M}{q^n}
		\sum_{\mbv\in\cV_q^n}
		\lambda(\mbv)
		\omega^{\langle\mbu,\mbv\rangle}
		W_\mbv.
	\end{align*}
	Taking the Hilbert--Schmidt inner product with \(P\) gives
	\begin{align*}
		\frac{1}{M}\Tr(W_\mbu P W_\mbu^\dagger P)
		=
		\frac{M}{q^n}
		\sum_{\mbv\in\cV_q^n}
		\omega^{\langle\mbu,\mbv\rangle}
		|\lambda(\mbv)|^2,
	\end{align*}
	which proves~\eqref{eq:qudit_pps_MI_coeff}, and hence~\eqref{eq:qudit_pps_MI}.
	
	For the pointwise inequality, we adopt the argument in~\cite{AL99}. Since \(P=P^\dagger=P^2\),
	\begin{align*}
		|\Tr(W_\mbv^\dagger P)|^2
		&=
		|\Tr(PW_\mbv^\dagger P)|^2\\
		&\le
		\Tr(P)\Tr(PW_\mbv P W_\mbv^\dagger P)\\
		&=
		M\Tr(W_\mbv P W_\mbv^\dagger P),
	\end{align*}
	where the inequality follows from the Cauchy--Schwarz inequality. Dividing by \(M^2\) gives
	\(\widehat\Lambda(\mbv)\ge\Lambda(\mbv)\), while
	\(\Lambda(\mbv)\ge0\) follows directly from its definition.
\end{proof}

\subsection{Shor--Laflamme weight enumerators}

The conventional Hamming-weight MacWilliams identity~\cite{SL97} is recovered by applying a Hamming-weight functional to the Weyl power spectrum and Weyl-conjugation autocorrelation.

Define the Hamming-weight functional on \(\cH_{\cV_q}\) by
\begin{align}
	\varphi_H(x,y)
	=
	x\bra{0}
	+
	y\sum_{v\in\cV_q\setminus\{0\}}\bra{v}.
	\label{eq:qudit_hamming_weight_functional}
\end{align}
Applying \(\varphi_H^{\otimes n}\) to the two formal vectors gives
\begin{align*}
	A(x,y)
	&=
	\varphi_H(x,y)^{\otimes n}\ket{\Lambda}
	=
	\sum_{j=0}^n A_jx^{n-j}y^j,\\
	B(x,y)
	&=
	\varphi_H(x,y)^{\otimes n}\ket{\widehat\Lambda}
	=
	\sum_{j=0}^n B_jx^{n-j}y^j,
\end{align*}
where
\begin{align*}
	A_j
	&=
	\sum_{\substack{\mbv\in\cV_q^n\\ \wt(\mbv)=j}}
	\Lambda(\mbv),
	&
	B_j
	&=
	\sum_{\substack{\mbv\in\cV_q^n\\ \wt(\mbv)=j}}
	\widehat\Lambda(\mbv).
\end{align*}
These are the Shor--Laflamme weight enumerators of the code~\cite{SL97}.

\begin{corollary}[Hamming-weight MacWilliams identity]
	\label{cor:qudit_HMI}
	The Hamming-weight enumerators \(A(x,y)\) and \(B(x,y)\) satisfy
	\begin{align}
		B(x,y)
		=
		\frac{M}{q^n}
		A\bigl(x+(q^2-1)y,x-y\bigr).
		\label{eq:qudit_hamming_MI}
	\end{align}
	Moreover,
	\begin{align}
		B_j\ge A_j\ge0,
		\qquad
		A_j=B_j\quad\text{for }0\le j<d.
		\label{eq:qudit_SL_constraints}
	\end{align}
\end{corollary}

\begin{proof}
	By \eqref{eq:qudit_fourier_operator},
	\begin{align*}
		\varphi_H\cF_q
		={}&
		\bigl(x+(q^2-1)y\bigr)\bra{0}
		+
		(x-y)
		\sum_{v\in\cV_q\setminus\{0\}}\bra{v}.
	\end{align*}
By character orthogonality~\eqref{eq:character_orthogonality},
	\(
		\sum_{u\in\cV_q\setminus\{0\}}
		\omega^{\langle u,v\rangle}
		=
		-1
	\)
	for \(v\neq0\), 
	whereas for \(v=0\), \(  \sum_{u\in\cV_q\setminus\{0\}}
	\omega^{\langle u,v\rangle}
	= q^2-1\).
	
	Since
	\(\varphi_H^{\otimes n}\cF_q^{\otimes n}
	=(\varphi_H\cF_q)^{\otimes n}\),
	applying \(\varphi_H^{\otimes n}\) to the Weyl power-spectrum MacWilliams identity~\eqref{eq:qudit_pps_MI} gives
\eqref{eq:qudit_hamming_MI}
	as claimed.
	
	The inequalities and equalities in~\eqref{eq:qudit_SL_constraints} follow from the pointwise inequality~\eqref{eq:qudit_pointwise_AB} and the quantum error correction condition~\eqref{eq:QEC_condition}.

\end{proof}

Comparing coefficients in~\eqref{eq:qudit_hamming_MI} gives
\begin{align*}
	B_j
	=
	\frac{M}{q^n}\sum_{r=0}^n
	K_j^{(q^2)}(r;n)A_r,
\end{align*}
where
\begin{align}
	K_j^{(q')}(r;n)
	=
	\sum_{s=0}^j
	(-1)^s(q'-1)^{j-s}
	\binom{r}{s}\binom{n-r}{j-s}  \label{eq:krawtchouk}
\end{align}
is the \(q'\)-ary Krawtchouk polynomial~\cite{MS77}.

\subsection{Complete Weyl enumerators}

The complete-weight MacWilliams identity~\cite{LHL16,hu2019complete} can be similarly recovered by applying a complete-weight functional to the Weyl power-spectrum identity.

Let \(\{z_v:v\in\cV_q\}\) be a set of formal variables, and define the complete-weight functional on \(\cH_{\cV_q}\) by
\begin{align*}
	\varphi_{\mathrm{cw}}
	=
	\sum_{v\in\cV_q}z_v\bra{v}.
\end{align*}
Applying \(\varphi_{\mathrm{cw}}^{\otimes n}\) to the two formal vectors gives the complete Weyl enumerators
\begin{align*}
	A_{\mathrm{cw}}(\{z_v\})
	&=
	\varphi_{\mathrm{cw}}^{\otimes n}\ket{\Lambda}
	=
	\sum_{\mbv\in\cV_q^n}
	\Lambda(\mbv)
	\prod_{j=1}^n z_{v_j},\\
	B_{\mathrm{cw}}(\{z_v\})
	&=
	\varphi_{\mathrm{cw}}^{\otimes n}\ket{\widehat\Lambda}
	=
	\sum_{\mbv\in\cV_q^n}
	\widehat\Lambda(\mbv)
	\prod_{j=1}^n z_{v_j}.
\end{align*}

\begin{corollary}[Complete-weight MacWilliams identity]
	\label{cor:qudit_CWE_MI}
	The complete Weyl enumerators satisfy
	\begin{align}
		B_{\mathrm{cw}}(\{z_v\})
		=
		\frac{M}{q^n}
		A_{\mathrm{cw}}(\{\widehat z_v\}),
		\label{eq:qudit_CWE_MI}
	\end{align}
	where
 		\(\widehat z_v
		=
		\sum_{u\in\cV_q}
		\omega^{\langle u,v\rangle}z_u,
		\)
		for 		$v\in\cV_q.$

\end{corollary}

\begin{proof}
	By~\eqref{eq:qudit_fourier_operator},
	\begin{align*}
		\varphi_{\mathrm{cw}}\cF_q
		&=
		\sum_{v\in\cV_q}
		\left(
		\sum_{u\in\cV_q}
		\omega^{\langle u,v\rangle}z_u
		\right)
		\bra{v}\notag\\
		&=
		\sum_{v\in\cV_q}\widehat z_v\bra{v}.
	\end{align*}
	Since
	\(\varphi_{\mathrm{cw}}^{\otimes n}\cF_q^{\otimes n}
	=(\varphi_{\mathrm{cw}}\cF_q)^{\otimes n}\),
	applying \(\varphi_{\mathrm{cw}}^{\otimes n}\) to the Weyl power-spectrum MacWilliams identity~\eqref{eq:qudit_pps_MI} gives \eqref{eq:qudit_CWE_MI}.
	
\end{proof}

\section{Twisted Fourier Transforms over Qudits}
\label{sec:qudit_twists}

The Weyl power-spectrum MacWilliams identity corresponds to the ordinary Fourier transform of the power spectrum. In the binary setting, the shadow transform is obtained by inserting a local sign that is \(+1\) on the identity and \(-1\) on the three nonidentity Pauli operators before applying the Fourier transform~\cite{Rains99b}. This motivates the following generalization of twisted Fourier transform.

Consider the normalized universal inversion map~\cite{RBC+01}
\begin{align}
	\mathcal I_q(A)
	=
	\frac{\Tr(A)I-A}{q-1},
	\label{eq:normalized_universal_inverter}
\end{align}
which acts on the Weyl basis as
\(\mathcal I_q(I)=I\) and
\(\mathcal I_q(W_v)=-\frac{1}{q-1}W_v\) for \(v\neq0\).

For a pure state \(\rho=\ket{\psi}\bra{\psi}\), the normalized universal inversion gives
\(
	\mathcal I_q(\rho)
	=
	\frac{I-\ket{\psi}\bra{\psi}}{q-1},
\)
which is the maximally mixed state on the subspace orthogonal to \(\ket{\psi}\). For \(q=2\), this subspace is one-dimensional and \(\mathcal I_2\) reduces to \(A\mapsto YA^T Y\), which underlies the Rains shadow transform~\cite{Rains99b}.

\begin{definition}[Qudit twists]
	\label{def:qudit_twists}
	For each \(t\in\cV_q\), we define two twists, denoted by \(t^+\) and \(t^-\). The \(t^+\) twist is generated by the Weyl-conjugation phase associated with \(W_t\) in~\eqref{eq:Weyl_conjugation}, whereas the \(t^-\) twist additionally applies the normalized universal inversion~\eqref{eq:normalized_universal_inverter}. Their actions on the formal Weyl space are defined by
	\begin{align}
		\Upsilon_{t^+}\ket{v}
		&=
		\omega^{\langle t,v\rangle}\ket{v},
		\label{eq:qudit_plus_twist}\\
		\Upsilon_{t^-}\ket{v}
		&=
		\begin{cases}
			\ket{0}, & v=0,\\[1mm]
			-\dfrac{1}{q-1}\,
			\omega^{\langle t,v\rangle}\ket{v},
			& v\neq0.
		\end{cases}
		\label{eq:qudit_minus_twist}
	\end{align}
	
	The corresponding twisted Fourier operators are
	\begin{align}
		\cT_{t^\pm}
		=
		\cF_q\Upsilon_{t^\pm}.
		\label{eq:TFU}
	\end{align}
	
\end{definition}
Note that $\cT_{0^+}= \cF_q$.

The phase \(\omega^{\langle t,v\rangle}\) in Definition~\ref{def:qudit_twists} is precisely the phase acquired by \(W_v\) under conjugation by \(W_t\) in~\eqref{eq:Weyl_conjugation}.

For \(q=2\), the two twists with \(t=0\) recover the familiar quantum weight-enumerator transforms. The \(0^+\) twist is trivial and gives the ordinary Shor--Laflamme transform~\cite{SL97}, whereas the \(0^-\) twist changes the signs of all three nonidentity Pauli operators and gives the Rains shadow transform~\cite{Rains99b}.  For each nonzero \(t\), the \(t^-\) twist changes the sign of \(W_t\) only, whereas the \(t^+\) twist changes the signs of the other two nonidentity Pauli operators.

Let \(\mbt=(t,\ldots,t)\in\cV_q^n\).

\begin{definition}[Twisted Weyl-conjugation autocorrelation]
	\label{def:twisted_autocorrelation}
	For an \(n\)-qudit quantum code with projector \(P\) of rank \(M\), define the twisted Weyl-conjugation autocorrelations by
	\begin{align}
		\widetilde\Lambda_{t^+}(\mbh)
		&=
		\frac{1}{M}
		\Tr\!\left(
		W_{\mbh+\mbt}P
		W_{\mbh+\mbt}^\dagger P
		\right),
		\label{eq:plus_twisted_autocorrelation}\\
		\widetilde\Lambda_{t^-}(\mbh)
		&=
		\frac{1}{M}
		\Tr\!\left(
		W_{\mbh+\mbt}
		\mathcal I_q^{\otimes n}(P)
		W_{\mbh+\mbt}^\dagger P
		\right),
		\label{eq:minus_twisted_autocorrelation}
	\end{align}
	for \(\mbh\in\cV_q^n\). The corresponding formal vectors are
	\begin{align}
		\ket{\widetilde\Lambda_{t^\pm}}
		=
		\sum_{\mbh\in\cV_q^n}
		\widetilde\Lambda_{t^\pm}(\mbh)\ket{\mbh}.
	\end{align}
\end{definition}

The \(0^+\) autocorrelation is the ordinary Weyl-conjugation autocorrelation \(\widehat\Lambda\), while the \(0^-\) autocorrelation is the qudit analogue of the Rains shadow autocorrelation~\cite{HESG18}.

\begin{theorem}[Twisted MacWilliams identity]
	\label{thm:qudit_twisted_MI}
	For every \(t\in\cV_q\),
	\begin{align}
		\ket{\widetilde\Lambda_{t^\pm}}
		=
		\frac{M}{q^n}
		\cT_{t^\pm}^{\otimes n}\ket{\Lambda}.
		\label{eq:qudit_twisted_MI}
	\end{align}
	Moreover,
		\begin{align}
		\widetilde\Lambda_{t^\pm}(\mbh)\ge0.
		\label{eq:twisted_spectrum_positivity}
	\end{align}
\end{theorem}

\begin{proof}
	We first consider the \(t^+\) twist. By~\eqref{eq:plus_twisted_autocorrelation}, we have the translation relation
	\begin{align}
		\widetilde\Lambda_{t^+}(\mbh)
		=
		\widehat\Lambda(\mbh+\mbt)
		\geq 0,
		\label{eq:translation}
	\end{align}
	where the inequality follows from~\eqref{eq:qudit_pointwise_AB}. Thus, the positivity of \(\widetilde\Lambda_{t^+}(\mbh)\) follows immediately.

	Applying the Weyl power-spectrum MacWilliams identity~\eqref{eq:qudit_pps_MI_coeff} gives
	\begin{align*}
		\widetilde\Lambda_{t^+}(\mbh)
		&=
		\frac{M}{q^n}
		\sum_{\mbv\in\cV_q^n}
		\omega^{\langle\mbh+\mbt,\mbv\rangle}
		\Lambda(\mbv)
		\nonumber\\
		&=
		\frac{M}{q^n}
		\sum_{\mbv\in\cV_q^n}
		\omega^{\langle\mbh,\mbv\rangle}
		\omega^{\langle\mbt,\mbv\rangle}
		\Lambda(\mbv),
	\end{align*}
	which is precisely the \(\mbh\)-component of
	\(\frac{M}{q^n}\cT_{t^+}^{\otimes n}\ket{\Lambda}\).
	
	For the \(t^-\) twist,  by the Weyl expansion~\eqref{eq:qudit_projector_expansion}, we have
	\begin{align*}
		\mathcal I_q^{\otimes n}(P)
		=
		\frac{M}{q^n}
		\sum_{\mbv\in\cV_q^n}
		\lambda(\mbv)
		\left(-\frac{1}{q-1}\right)^{\wt(\mbv)}
		W_\mbv.
	\end{align*}
	Taking the Weyl conjugation in~\eqref{eq:Weyl_conjugation} gives
	\begin{align*}
		&
		W_{\mbh+\mbt}
		\mathcal I_q^{\otimes n}(P)
		W_{\mbh+\mbt}^\dagger
		\nonumber\\
		&\qquad=
		\frac{M}{q^n}
		\sum_{\mbv\in\cV_q^n}
		\lambda(\mbv)
		\left(-\frac{1}{q-1}\right)^{\wt(\mbv)}
		\omega^{\langle\mbh+\mbt,\mbv\rangle}
		W_\mbv.
	\end{align*}
	Since $\Tr(W_\mbv P)
		=
		M\lambda(\mbv)^*,$
	the above equation leads to
	\begin{align*}
		\widetilde\Lambda_{t^-}(\mbh)
		=
		\frac{M}{q^n}
		\sum_{\mbv\in\cV_q^n}
		\left(-\frac{1}{q-1}\right)^{\wt(\mbv)}
		\omega^{\langle\mbh+\mbt,\mbv\rangle}
		\Lambda(\mbv),
	\end{align*}
which, by a direct calculation, is the \(\mbh\)-component of
\(\frac{M}{q^n}\cT_{t^-}^{\otimes n}\ket{\Lambda}\).

For the positivity, fix an ordering of the computational basis and let \(L_{ab}=\ket{a}\bra{b}-\ket{b}\bra{a}\) for \(a<b\). The universal inversion can be written as
\begin{align*}
	\mathcal I_q(A)
	=
	\frac{1}{q-1}
	\sum_{a<b}
	L_{ab}A^T L_{ab}^\dagger.
\end{align*}
Since the tensor product of the local transpositions is the full transpose,
\begin{align*}
	\mathcal I_q^{\otimes n}(P)
	=
	\frac{1}{(q-1)^n}
	\sum_{\substack{a_1<b_1,\ldots,a_n<b_n}}
	L_{\mathbf a\mathbf b}
	P^T
	L_{\mathbf a\mathbf b}^\dagger
	\succeq0,
\end{align*}
where \(L_{\mathbf a\mathbf b}=L_{a_1b_1}\otimes\cdots\otimes L_{a_nb_n}\), and the inequality follows from \(P^T\succeq0\). Therefore, by~\eqref{eq:minus_twisted_autocorrelation},
\begin{align*}
	\widetilde\Lambda_{t^-}(\mbh)
	=
	\frac{1}{M}
	\Tr\!\left[
	W_{\mbh+\mbt}
	\mathcal I_q^{\otimes n}(P)
	W_{\mbh+\mbt}^\dagger P
	\right]
	\ge0,
\end{align*}
since both \(W_{\mbh+\mbt}\mathcal I_q^{\otimes n}(P)W_{\mbh+\mbt}^\dagger\) and \(P\) are positive semidefinite.

\end{proof}

\subsection{Binary specialization: the eight twists}
\label{subsec:binary_twists}

For \(q=2\), the normalized universal inversion acts on the Pauli operators as
\begin{align*}
	&\mathcal I_2(I)=I,
	\qquad
	\mathcal I_2(X)=-X,\\
	&\mathcal I_2(Y)=-Y,
	\qquad
	\mathcal I_2(Z)=-Z.
\end{align*}
Equivalently, 
\begin{align}
	\mathcal I_2(A)=YA^TY \label{eq:I_2}
\end{align}
for $A\in\{I,X,Y,Z\}$.
Moreover, the Weyl-conjugation phases reduce to signs.

Let \(\tau=\tau_X\tau_Y\tau_Z\in\{0,1\}^3\) denote the binary sign label, where \(\tau_W=1\) indicates a sign change of the Pauli operator \(W\). The \(2q^2=8\) qudit twists reduce to the eight binary sign twists, with the following correspondence:
\begin{align}
	\begin{array}{c|c}
		\text{qudit twist}&\text{binary sign label}\\
		\hline
		0^+&000\\
		X^+&011\\
		Z^+&110\\
		Y^+&101\\
		0^-&111\\
		X^-&100\\
		Z^-&001\\
		Y^-&010
	\end{array}
	\label{eq:binary_twist_correspondence}
\end{align}
In particular, \(0^+\leftrightarrow000\) recovers the ordinary Shor--Laflamme MacWilliams transform~\cite{SL97}, while \(0^-\leftrightarrow111\) recovers the Rains shadow transform~\cite{Rains99b}. The remaining six twists, corresponding to nonzero \(t\in\cV_2\), are their Pauli translations by Theorem~\ref{thm:two_class_reduction}. For each nonzero \(t\), the \(t^-\) twist changes the sign of \(W_t\) only, whereas the \(t^+\) twist changes the signs of the other two nonidentity Pauli operators.

\subsection{Bell sampling interpretation of the binary twists}
\label{subsec:bell_101}

Bell sampling provides a natural operational interpretation of the binary twisted autocorrelations. For pure states, conjugate-copy Bell sampling has previously been used to sample the Pauli probability distribution~\cite{LC22}. We first extend this connection to quantum-code projectors and show that conjugate-copy and identical-copy Bell sampling correspond to two distinguished binary twists. 

The binary twisted autocorrelations admit a direct interpretation in terms of two-copy Bell sampling. Define the \(n\)-qubit Bell basis by
\begin{align}
	\ket{\Phi_\mbh}
	=
	(I\otimes W_\mbh)\ket{\Phi^+}^{\otimes n}
\end{align}
for \(\mbh\in\cV_2^n\), where
\(\ket{\Phi^+}=\frac{1}{\sqrt{2}}(\ket{00}+\ket{11})\).

Let \(\rho=P/M\). We consider two Bell-sampling distributions. Let \(p_+(\mbh)\) denote the probability of obtaining the Bell outcome \(\mbh\) from the conjugate-copy state \(\rho^*\otimes\rho\), and let \(p_-(\mbh)\) denote the corresponding probability from the identical-copy state \(\rho\otimes\rho\):
\begin{align*}
	p_+(\mbh)
	&=
	\bra{\Phi_\mbh}
	\rho^*\otimes\rho
	\ket{\Phi_\mbh},\\
	p_-(\mbh)
	&=
	\bra{\Phi_\mbh}
	\rho\otimes\rho
	\ket{\Phi_\mbh}.
\end{align*}

\begin{theorem}[Bell-sampling interpretation of the binary twists]
	\label{thm:bell_sampling_binary_twists}
	For every \(\mbh\in\cV_2^n\),
	\begin{align}
		p_+(\mbh)
		&=
		\frac{1}{2^nM}
		\widehat\Lambda(\mbh)
		=
		\frac{1}{2^nM}
		\widetilde\Lambda_{0^+}(\mbh),
		\label{eq:bell_sampling_0_plus}\\
		p_-(\mbh)
		&=
		\frac{1}{2^nM}
		\widetilde\Lambda_{Y^-}(\mbh).
		\label{eq:bell_sampling_Y_minus}
	\end{align}
	Thus, conjugate-copy Bell sampling realizes the normalized \(0^+=000\) autocorrelation, whereas identical-copy Bell sampling realizes the normalized \(Y^-=010\) twisted autocorrelation.
\end{theorem}

\begin{proof}
\begin{align*}
	p_+(\mbh)
	&=
	\bra{\Phi_\mbh}
	\rho^*\otimes\rho
	\ket{\Phi_\mbh}\\
	&=
	\bra{\Phi^+}^{\otimes n}
	\rho^*
	\otimes
	W_\mbh^\dagger\rho W_\mbh
	\ket{\Phi^+}^{\otimes n}\\
	&\stackrel{(a)}{=}
	\frac{1}{2^n}
	\Tr\!\left(
	(\rho^*)^T
	W_\mbh^\dagger\rho W_\mbh
	\right)\\
	&\stackrel{(b)}{=}
	\frac{1}{2^nM^2}
	\Tr\!\left(
	W_\mbh P W_\mbh^\dagger P
	\right)
	=
	\frac{1}{2^nM}
	\widehat\Lambda(\mbh),
\end{align*}
where \((a)\) follows from the identity
\begin{align*}
	\bra{\Phi^+}^{\otimes n}
	A\otimes B
	\ket{\Phi^+}^{\otimes n}
	=
	\frac{1}{2^n}\Tr(A^TB),
\end{align*}
and \((b)\) follows from \((\rho^*)^T=\rho=P/M\) and the cyclicity of the trace. Thus, \(p_+(\mbh)\) is precisely the normalized \(0^+\) Weyl-conjugation autocorrelation of \(P\).

 Let \(\mby=((1,1),\ldots,(1,1))\in\cV_2^n\), corresponding to \(Y^{\otimes n}\) up to an overall phase.  Then
\begin{align*}
	p_-(\mbh)
	&=
	\bra{\Phi_\mbh}
	\rho\otimes\rho
	\ket{\Phi_\mbh}\\
	&=
	\frac{1}{2^n}
	\Tr\!\left(
	\rho^T W_\mbh^\dagger\rho W_\mbh
	\right)\\
	&\stackrel{(c)}{=}
	\frac{1}{2^nM^2}
	\Tr\!\left(
	W_{\mbh+\mby}
	\mathcal I_2^{\otimes n}(P)
	W_{\mbh+\mby}^\dagger P
	\right)\\
	&=
	\frac{1}{2^nM}
	\widetilde\Lambda_{Y^-}(\mbh),
\end{align*}
where $(c)$ follows from \eqref{eq:I_2}.

\end{proof}

\begin{figure*}[htbp]
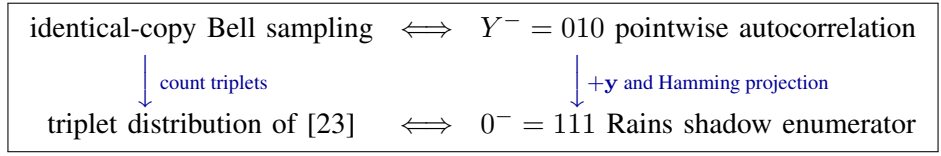

	\centering
	\[
	\boxed{
		\begin{array}{ccc}
			\text{identical-copy Bell sampling}
			&\Longleftrightarrow&
			Y^-=010\text{ pointwise autocorrelation}
			\\[1mm]
			{\color{blue!60!black}\Big\downarrow\scriptstyle{\text{ count triplets}}}
			&&
			{\color{blue!60!black}\Big\downarrow\scriptstyle{+\mby\text{ and Hamming projection}}}
			\\[1mm]
			\text{triplet distribution of~\cite{MLP+26}}
			&\Longleftrightarrow&
			0^-=111\text{ Rains shadow enumerator}
	\end{array}}
	\]
	\caption{Relation between the pointwise \(Y^-\) twisted autocorrelation and the Bell-sampling interpretation of the Rains shadow enumerator in~\cite{MLP+26}.}
	\label{fig:bell_shadow_relation}
\end{figure*}

\begin{corollary}[Bell-sampling realization of the binary twists]
	\label{cor:bell_sampling_binary_twists}
	For every \(t\in\cV_2\), let \(\mbt=(t,\ldots,t)\in\cV_2^n\). Then
	\begin{align}
		\frac{1}{2^nM}
		\widetilde\Lambda_{t^+}(\mbh)
		&=
		p_+(\mbh+\mbt),
		\label{eq:bell_sampling_plus_twists}\\
		\frac{1}{2^nM}
		\widetilde\Lambda_{t^-}(\mbh)
		&=
		p_-(\mbh+\mbt+\mby).
		\label{eq:bell_sampling_minus_twists}
	\end{align}
	Consequently, all eight normalized binary twisted autocorrelations are realized by conjugate-copy or identical-copy Bell sampling, up to fixed translations of the Bell-outcome labels.
\end{corollary}

\begin{proof}
	The result follows directly from the translation relation~\eqref{eq:translation} and the definitions of \(p_+\) and \(p_-\).
\end{proof}

The relation to the Bell-sampling interpretation of shadow enumerators in~\cite{MLP+26} follows by coarse-graining the pointwise distribution in~\eqref{eq:bell_sampling_Y_minus}. In the Bell basis, \(Y\) (\(11\)) labels the singlet outcome, whereas \(I,X,Z\) (\(00,10,01\)) label the three triplet outcomes. Hence the number of triplet outcomes in \(\mbh\) is \(\wt(\mbh+\mby)\). Moreover, the translation relation gives
\begin{align*}
	\widetilde\Lambda_{Y^-}(\mbh)
	=
	\widetilde\Lambda_{0^-}(\mbh+\mby).
\end{align*}
Therefore, summing \(p_-(\mbh)\) over all Bell outcomes containing exactly \(j\) triplets is precisely the Hamming-weight-\(j\) projection of the $0^-=111$ twisted autocorrelation, recovering the Rains shadow-enumerator interpretation of~\cite{MLP+26}. Figure~\ref{fig:bell_shadow_relation} summarizes this relation.

The parallelized SWAP-test interpretation in~\cite{SGZZ26} is equivalent to the same local coarse-graining: the three triplet Bell outcomes \(I,X,Z\) are combined into the symmetric outcome, while the singlet outcome \(Y\) gives the antisymmetric outcome, followed by the same weight projection.

%
%

\section{Redundancy of Twisted Fourier Constraints}
\label{sec:twist_redundancy}

The preceding sections establish a unified Fourier framework
for the Shor--Laflamme and shadow identities and their twisted
generalizations. We now examine whether the twisted identities
yield independent positivity constraints for LP bounds.
We first establish their translation structure and then study
which constraints remain distinct after applying linear functionals.

\subsection{Translation structure of twisted Fourier transforms}
\label{subsec:two_class_reduction}

The twisted Fourier transforms differ from the ordinary and
shadow transforms only by translations of their output coefficients.

\begin{theorem}[Translation identity for twisted Fourier transforms]
	\label{thm:two_class_reduction}
	For $t\in\cV_q$, define the permutation operator
	\begin{align}
		\cP_t\ket{h}=\ket{h-t}.
	\end{align}
	Then
	\begin{align}
		\cT_{t^\pm}=\cP_t\cT_{0^\pm}.
		\label{eq:two_class_operator_identity}
	\end{align}
	Consequently, for arbitrary labels
	$\mbt=(t_1,\ldots,t_n)\in\cV_q^n$,
	\begin{align}
		\bigotimes_{j=1}^n\cT_{t_j^\pm}
		=\cP_{\mbt}\cT_{0^\pm}^{\otimes n},
		\label{eq:two_class_blockwise_identity}
	\end{align}
	where $\cP_{\mbt}=\bigotimes_{j=1}^n\cP_{t_j}$
	and the sign $\pm$ is uniform across all coordinates.
\end{theorem}

\begin{proof}
	By bilinearity of the symplectic form,
	\begin{align*}
		(\cF_q\Upsilon_{t^+})_{h,v}
		&=\omega^{\langle h+t,v\rangle}
		=(\cP_t\cF_q)_{h,v}.
	\end{align*}
	Since $\Upsilon_{t^-}=\Upsilon_{t^+}\Upsilon_{0^-}$,
	the same identity holds for the minus twist.
	Taking tensor products completes the proof.
\end{proof}

The transforms $0^+$ and $0^-$ correspond to the ordinary
MacWilliams and Rains shadow transforms, respectively.
For a vector in the Weyl basis, inequalities are understood
componentwise.

\begin{proposition}[Positivity equivalence of twisted Fourier transforms]
	\label{prop:twist_positivity_equivalence}
	Let $\ket{a}$ be a Weyl-resolved coefficient vector,
	and let $\mbt\in\cV_q^n$.
	For either uniform sign $\pm$, the vectors
	\begin{align*}
		\left(\bigotimes_{j=1}^n\cT_{t_j^\pm}\right)\ket{a}
		\quad\text{and}\quad
		\cT_{0^\pm}^{\otimes n}\ket{a}
	\end{align*}
	have the same coefficients up to permutation.
	Consequently, whenever these coefficients are real,
	\begin{align*}
		\left(\bigotimes_{j=1}^n\cT_{t_j^\pm}\right)
		\ket{a}\ge0
		\quad\Longleftrightarrow\quad
		\cT_{0^\pm}^{\otimes n}\ket{a}\ge0.
	\end{align*}
\end{proposition}

\begin{proof}
	By \eqref{eq:two_class_blockwise_identity},
	$
	\left(\bigotimes_{j=1}^n\cT_{t_j^\pm}\right)
	\ket{a}
	=
	\cP_{\mbt}
	\cT_{0^\pm}^{\otimes n}\ket{a}.$
	Since $\cP_{\mbt}$ is a permutation operator,
	the two vectors contain the same coefficients
	and satisfy equivalent componentwise
	nonnegativity conditions.
\end{proof}

\subsection{Projected Twisted Fourier Transforms}
\label{subsec:projected_twists}

The exact Weyl power spectrum contains $q^{2n}$
coefficients, making direct LP
calculations impractical. We therefore consider
linear functionals that aggregate Weyl labels
into a smaller number of variables.

Let $\Pi=\{C_0,C_1,\ldots,C_r\}$ be a partition
of $\cV_q$, with $C_0=\{0\}$.
For formal variables $x_0,\ldots,x_r$, define
\begin{align}
	\ell_\Pi(x)
	=\sum_{i=0}^r x_i
	\sum_{v\in C_i}\bra{v}.
	\label{eq:partition_functional}
\end{align}
For any Weyl-resolved coefficient vector
$\ket{a}=\sum_v a_v\ket{v}$,
$\ell_\Pi(x)\ket{a}
	=\sum_{i=0}^r x_i
	\sum_{v\in C_i}a_v.$

We say that a transform $\cT$ is
\emph{closed under $\Pi$} if
\begin{align}
	\ell_\Pi(x_0,\ldots,x_r)\cT
	=\ell_\Pi(y_0,\ldots,y_r),
\end{align}
where each $y_i$ is a linear form in
$x_0,\ldots,x_r$.
Thus, the transformed functional depends
only on the same projected variables.

\begin{theorem}[Redundancy of closed twisted Fourier constraints]
	\label{thm:projected_twist_no_go}
	Suppose that $\cF_q$ is closed under $\Pi$.
	If $\cT_{t^\pm}$ is also closed under $\Pi$,
	then its projected output coefficients
	are permutations of those of $\cT_{0^\pm}$.
	Consequently, the corresponding
	coefficientwise positivity constraints
	are equivalent.
\end{theorem}

\begin{proof}
	Suppose first that $\cT_{t^+}$ is closed under $\Pi$.
	Since both $\cF_q$ and $\cT_{t^+}$ are closed
	under $\Pi$, there exist linear forms
	$y_i$ and $z_i$ in $x_0,\ldots,x_r$ such that
	\begin{align*}
		\ell_\Pi(x_0,\ldots,x_r)\cF_q
		&=\ell_\Pi(y_0,\ldots,y_r),\\
		\ell_\Pi(x_0,\ldots,x_r)\cT_{t^+}
		&=\ell_\Pi(z_0,\ldots,z_r).
	\end{align*}
	Since $\cF_q$ is invertible, its restriction
	to the space spanned by the partition functionals
	is also invertible. Thus, the substitution
	$(x_0,\ldots,x_r)\mapsto(y_0,\ldots,y_r)$
	is invertible. Moreover,
	\begin{align*}
		\ell_\Pi(x_0,\ldots,x_r)
		&=\ell_\Pi(x_0,\ldots,x_r)
		\cF_q\cF_q^{-1}\\
		&=\ell_\Pi(y_0,\ldots,y_r)
		\cF_q^{-1}.
	\end{align*}
	Hence $\cF_q^{-1}$ is also closed under $\Pi$.
	
	By the translation identity
	\eqref{eq:two_class_operator_identity},
	\begin{align*}
		\ell_\Pi(x_0,\ldots,x_r)\cP_t
		&=\ell_\Pi(x_0,\ldots,x_r)
		\cT_{t^+}\cF_q^{-1}\\
		&=\ell_\Pi(z_0,\ldots,z_r)
		\cF_q^{-1}\\
		&=\ell_\Pi(w_0,\ldots,w_r),
	\end{align*}
	where each $w_i$ is a linear form in
	$x_0,\ldots,x_r$.
	Thus, $\cP_t$ is closed under $\Pi$.
	
	On the other hand,
	\begin{align*}
		\ell_\Pi(x_0,\ldots,x_r)\cP_t
		=\sum_{i=0}^r x_i
		\sum_{v\in C_i}\bra{v+t}.
	\end{align*}
	Closure requires the coefficients of
	$\bra{v}$ to be identical for all
	$v$ in the same cell $C_j$.
	Since $x_0,\ldots,x_r$ are independent
	formal variables, $C_j-t$ must be
	contained in a single cell of $\Pi$.
	Furthermore, translation by $-t$
	has finite order, so these inclusions
	force translation to permute the cells.
Hence there exists a permutation $\mu$
of $\{0,\ldots,r\}$ such that
$C_j-t=C_{\mu(j)}$.

Consequently,
\begin{align*}
	\ell_\Pi(x_0,\ldots,x_r)\cP_t
	&=\ell_\Pi(x_{\mu(0)},\ldots,
	x_{\mu(r)}),
\end{align*}
and therefore
\begin{align*}
	\ell_\Pi(x_0,\ldots,x_r)\cT_{t^+}
	&=\ell_\Pi(x_{\mu(0)},\ldots,
	x_{\mu(r)})\cT_{0^+}.
\end{align*}
Thus, the projected output coefficients
differ only by a permutation.

For the minus sector, recall from
\eqref{eq:TFU} that
$\cT_{t^-}=\cT_{t^+}\Upsilon_{0^-}$.
Since $C_0=\{0\}$, both $\Upsilon_{0^-}$
and its inverse are closed under $\Pi$
by \eqref{eq:qudit_minus_twist}.
Hence $\cT_{t^-}$ is closed under $\Pi$
if and only if $\cT_{t^+}$ is closed
under $\Pi$.
The same permutation identity follows
from $\cT_{t^-}=\cP_t\cT_{0^-}$.
\end{proof}

For the Hamming partition
$\Pi_H=\bigl\{\{0\},\cV_q\setminus\{0\}\bigr\}$,
only $t=0$ is $\Pi_H$-compatible: for $t\ne0$,
translation maps $\{0\}$ to $\{t\}$, which is
not a cell of $\Pi_H$.
The ordinary transform is given by
\eqref{eq:qudit_hamming_MI}, while the shadow transform gives
\begin{align}
	S(x,y)
	=\frac{M}{q^n}
	A\!\left(x+(q^2-1)y,\frac{y-x}{q-1}\right).
	\label{eq:qudit_shadow_hamming}
\end{align}
For $q=2$, this reduces to the conventional
Rains shadow identity
$S_2(x,y)=M2^{-n}A(x+3y,y-x)$.

Theorem~\ref{thm:projected_twist_no_go} gives a partial
no-go result: twisted transforms that close on the same
partition variables yield no new coefficientwise positivity
constraints. It does not preclude stronger LP constraints
from finer collections of linear functionals, as developed
in the next section.

\section{Refined Weight Enumerators and Linear Programming}
\label{sec:LP}

Although twisted Fourier identities do not yield additional
positivity constraints under compatible projections,
stronger constraints can be obtained by applying refined
functionals to both sides of these identities.
This leads to refined (split) weight enumerators and corresponding
MacWilliams and shadow identities, which provide stronger
LP constraints for stabilizer codes.

\subsection{Split Enumerators and Structural Linear Constraints}
\label{subsec:split_structural_lp}

We now focus on stabilizer codes~\cite{Got97,KKKS06}.
Let $\cS\subseteq\cV_q^n$ be an additive
stabilizer group and $\cS^\perp$ its
trace-symplectic dual. 
By \eqref{eq:qudit_power_spectrum}~and~\eqref{eq:qudit_dual_spectrum}, we have
\begin{align*}
	\Lambda(\mbv)
	&=\mathds{1}_{\cS}(\mbv),\\
	\widehat\Lambda(\mbv)
	&=\mathds{1}_{\cS^\perp}(\mbv).
\end{align*}

The two-class reduction does not preclude additional
linear constraints arising from finer weight enumerators~\cite{CRSS98,LA18}.
We therefore consider three levels of enumerator resolution:
$H$, $ H\times R_X$, and  $R_X\times R_Y.$
Here, $H$ denotes the Hamming weight enumerator, which
records only the total Hamming weight. The $H\times R_X$
and $R_X\times R_Y$ enumerators instead distinguish
the Weyl-operator compositions on selected coordinate
blocks, allowing us to incorporate additional
stabilizer-dependent constraints.

\paragraph{Split enumerators.}
Recall that $W_{a,b}=X(a)Z(b)$, where
$(a,b)\in\cV_q=\mathbb F_q^2$.

We introduce the $R_X$ refinement as a
Fourier-invariant partition of the Weyl
phase space, following the general framework
of~\cite{ZinEri96,gluesing2015fourier}.
Let
$L_X=\{(a,0):a\in\mathbb F_q\}$
and consider its affine cosets
$C_b=\{(a,b):a\in\mathbb F_q\}$.
The refinement distinguishes individual
points on $L_X$, while treating each
$C_b$, $b\neq0$, as a single cell.
Thus,
\begin{align*}
	R_X=\big\{
	&\{(a,0)\}:a\in\mathbb F_q,\\
	&\{(c,b):c\in\mathbb F_q\}:
	b\in\mathbb F_q^\times
	\big\},
\end{align*}
which consists of $2q-1$ cells.
To verify Fourier invariance, 
 consider the $(2q-1)$-dimensional formal vector space
$
	\cV_{R_X}
	=\operatorname{span}\left\{
	\ket{(a,0)},\ket{C_b}:
	a\in\mathbb F_q,\;
	b\in\mathbb F_q^\times
	\right\},
$
where 
$\ket{C_b}=\sum_{a\in\mathbb F_q}\ket{(a,b)}$.
By \eqref{eq:qudit_fourier_operator},
\begin{align*}
	\cF_q\ket{C_b}
	&=
	\sum_{a,c,d\in\mathbb F_q}
	\omega^{\operatorname{Tr}(da-cb)}
	\ket{(c,d)}\\
	&\stackrel{(a)}{=}
	\sum_{c,d\in\mathbb F_q}
	\omega^{-\operatorname{Tr}(cb)}
	q\delta_{d,0}\ket{(c,d)}\\
	&=
	q\sum_{c\in\mathbb F_q}
	\omega^{-\operatorname{Tr}(bc)}
	\ket{(c,0)},
\end{align*}
where $(a)$ follows the character orthogonality \eqref{eq:character_orthogonality}. Also,
\begin{align*}
	\cF_q\ket{(a,0)}
	&=
	\sum_{c,d\in\mathbb F_q}
	\omega^{\operatorname{Tr}(da)}
	\ket{(c,d)}\\
	&=
	\sum_{d\in\mathbb F_q}
	\omega^{\operatorname{Tr}(ad)}
	\ket{C_d}.
\end{align*}
Thus both transforms belong to $\cV_{R_X}$.
As $\cF_q$ is invertible, it follows that
$\cF_q\cV_{R_X}=\cV_{R_X}$.

The $R_Y$ refinement is defined analogously
using the diagonal direction
$L_Y=\{(a,a):a\in\mathbb F_q\}$:
\begin{align*}
	R_Y=\big\{
	&\{(a,a)\}:a\in\mathbb F_q,\\
	&\{(c,c+b):c\in\mathbb F_q\}:
	b\in\mathbb F_q^\times
	\big\}.
\end{align*}
Its associated cell subspace is likewise
invariant under $\cF_q$.
Both refinements consist of $2q-1$ cells.

The corresponding weight functionals are
\begin{align*}
	\varphi_{R_X}
	&=
	y_0\bra{0}
	+\sum_{a\neq0}y_a\bra{(a,0)}
	+\sum_{b\neq0}y_b^\perp
	\sum_{a\in\mathbb F_q}\bra{(a,b)},\\
	\varphi_{R_Y}
	&=
	z_0\bra{0}
	+\sum_{a\neq0}z_a\bra{(a,a)}
	+\sum_{b\neq0}z_b^\perp
	\sum_{a\in\mathbb F_q}\bra{(a,a+b)}.
\end{align*}

Partition the coordinates as
$T\sqcup U=[n]$, with $|T|=t$
and $|U|=u$.
Using the Hamming-weight functional
$\varphi_H$ defined in
\eqref{eq:qudit_hamming_weight_functional},
the full $H\times R_X$ split enumerators
are
\begin{align*}
	A_{H\times R_X}
	&=
	\left(
	\varphi_H(x_0,x_1)^{\otimes t}
	\otimes\varphi_{R_X}^{\otimes u}
	\right)|\Lambda\rangle\\
	&=
	\sum_{i,\mathbf j,\boldsymbol\ell}
	A_{i,\mathbf j,\boldsymbol\ell}
	x_0^{t-i}x_1^i
	y_0^{u-|\mathbf j|-|\boldsymbol\ell|}
	\prod_{a\neq0}y_a^{j_a}
	\prod_{b\neq0}(y_b^\perp)^{\ell_b},\\
	B_{H\times R_X}
	&=
	\left(
	\varphi_H(x_0,x_1)^{\otimes t}
	\otimes\varphi_{R_X}^{\otimes u}
	\right)|\widehat\Lambda\rangle\\
	&=
	\sum_{i,\mathbf j,\boldsymbol\ell}
	B_{i,\mathbf j,\boldsymbol\ell}
	x_0^{t-i}x_1^i
	y_0^{u-|\mathbf j|-|\boldsymbol\ell|}
	\prod_{a\neq0}y_a^{j_a}
	\prod_{b\neq0}(y_b^\perp)^{\ell_b}.
\end{align*}
Here $\mathbf j=(j_a)_{a\in\mathbb F_q^\times}$
and $\boldsymbol\ell=(\ell_b)_{b\in\mathbb F_q^\times}$
are nonnegative integer vectors, with
$|\mathbf j|=\sum_{a\neq0}j_a$ and
$|\boldsymbol\ell|=\sum_{b\neq0}\ell_b$.
The sum is over $0\le i\le t$ and
$|\mathbf j|+|\boldsymbol\ell|\le u$.

The $R_X\times R_Y$ split enumerators
are defined analogously by applying
$\varphi_{R_X}^{\otimes t}\otimes
\varphi_{R_Y}^{\otimes u}$ to
$|\Lambda\rangle$ and
$|\widehat\Lambda\rangle$.
Their coefficients are denoted by
$A_{\mathbf j,\boldsymbol\ell,
	\mathbf r,\mathbf s}$ and
$B_{\mathbf j,\boldsymbol\ell,
	\mathbf r,\mathbf s}$,
where
$\mathbf j,\boldsymbol\ell,
\mathbf r,\mathbf s
\in\mathbb Z_{\geq0}^{q-1}$
satisfy
$|\mathbf j|+|\boldsymbol\ell|\le t$
and
$|\mathbf r|+|\mathbf s|\le u$.
%
Applying these functionals to the
MacWilliams and shadow identities
yields the corresponding refined
identities, which are closed under
the Fourier transform.

\paragraph{Reference stabilizers}
Suppose that $\cS$ contains an element
of weight $u$.
After a coordinate permutation
and suitable local Clifford
transformations, we may choose
a reference stabilizer
\begin{align*}
	g=I^{\otimes t}\otimes X(1)^{\otimes u},
\end{align*}
where $t=n-u$.
We take $U=\operatorname{supp}(g)$
and $T=[n]\setminus U$.

Since $g\in\cS$, the $H\times R_X$
split enumerator satisfies
\begin{align*}
	A_{0,u\mathbf e_1,\mathbf0}\ge1,
\end{align*}
where $\mathbf e_1$ denotes the unit
vector indexed by $1\in\mathbb F_q^\times$.

Moreover, every
$\mbv=(\mba,\mbb)\in\cS^\perp$
must commute with $g$, implying
$
	\operatorname{Tr}_{\mathbb F_q/\mathbb F_p}
	\left(\sum_{r\in U}b_r\right)=0.
$
Since $\ell_b$ counts the coordinates
in the cell $C_b$ with $b\neq0$,
the $H\times R_X$ coefficients obey
\begin{align*}
	A_{i,\mathbf j,\boldsymbol\ell}
	=B_{i,\mathbf j,\boldsymbol\ell}=0
\end{align*}
whenever
$
	\operatorname{Tr}_{\mathbb F_q/\mathbb F_p}
	\left(
	\sum_{b\in\mathbb F_q^\times}
	b\ell_b
	\right)\neq0.
$

In particular, for $q=2$, there is
only one transverse cell, and the
condition reduces to
\begin{align*}
	A_{i,j,\ell}=B_{i,j,\ell}=0
	\qquad\text{if $\ell$ is odd}.
\end{align*}


\paragraph{Distance constraints}
Since $\cS^\perp\setminus\cS$
contains no Weyl vector of weight
less than $d$, the $H\times R_X$
split coefficients satisfy
\begin{align*}
	A_{i,\mathbf j,\boldsymbol\ell}
	=B_{i,\mathbf j,\boldsymbol\ell},
	\qquad
	i+|\mathbf j|+|\boldsymbol\ell|<d.
\end{align*}
Similarly, the $R_X\times R_Y$
split coefficients satisfy
\begin{align*}
	A_{\mathbf j,\boldsymbol\ell,
		\mathbf r,\mathbf s}
	=
	B_{\mathbf j,\boldsymbol\ell,
		\mathbf r,\mathbf s},
\end{align*}
whenever  $|\mathbf j|+|\boldsymbol\ell|
	+|\mathbf r|+|\mathbf s|<d.$

	\paragraph{Stabilizer orbits}
	For $q=2$, multiplication by the reference
	stabilizer $g=I^{\otimes t}\otimes X^{\otimes u}$
	exchanges $I\leftrightarrow X$ and
	$Y\leftrightarrow Z$ on $U$.
	Since $\cS$ and $\cS^\perp$ are invariant
	under this operation, we obtain
	\begin{align*}
		A_{i,j,\ell}
		&=A_{i,u-j-\ell,\ell},\\
		B_{i,j,\ell}
		&=B_{i,u-j-\ell,\ell}.
	\end{align*}

	For $u=2j+\ell$, multiplication by $g$
	preserves the composition $(i,j,\ell)$.
	Moreover, every Weyl vector is paired
	with a distinct vector obtained by
	multiplication by $g$.
	Since both vectors belong to the same
	composition class, its stabilizer and
	dual multiplicities must be even:
	\begin{align*}
		A_{i,j,\ell}
		\equiv B_{i,j,\ell}
		\equiv 0\pmod 2.
	\end{align*}

\subsection{Linear Programming Formulation}
\label{subsec:LP_formulation}

We now combine the projected MacWilliams
and shadow identities with the structural
constraints derived above to obtain
linear-programming bounds on stabilizer codes.

For an $[[n,k,d]]q$ stabilizer code,
the weight distributions $\{A_j\}_{j=0}^n$ and $\{B_j\}_{j=0}^n$ satisfy
\begin{align*}
	&A_0=B_0=1,\qquad
	0\le A_j\le B_j, \nonumber\\
	&A_j=B_j,\qquad 1\le j<d, \nonumber\\
	&\sum_j A_j=q^{n-k},\qquad
	\sum_j B_j=q^{n+k},
\end{align*}
together with the MacWilliams identity
\eqref{eq:qudit_hamming_MI}
and the shadow identity~\eqref{eq:qudit_shadow_hamming}.

For $q=2$, we additionally impose the three-type
counting constraints of~\cite[Theorem~16]{LA18}.
Specifically, one of the following three cases must hold:
\begin{align*}
	\text{I:}\quad&
	\sum_{j\text{ even}}A_j=2^{n-k-1},
	&
	\sum_{j\text{ even}}B_j=2^{n+k-1};\\
	\text{II:}\quad&
	\sum_{j\text{ even}}A_j=2^{n-k},
	&
	\sum_{j\text{ even}}B_j
	=2^{n-k-1}(4^k+2^k);\\
	\text{III:}\quad&
	\sum_{j\text{ even}}A_j=2^{n-k},
	&
	\sum_{j\text{ even}}B_j
	=2^{n-k-1}(4^k-2^k).
\end{align*}
An alternative proof of the general three-type
counting identities using quadratic Gauss sums
is provided in Appendix~\ref{app:three_type_counting}.

All applicable structural constraints discussed in
Section~\ref{subsec:split_structural_lp}
are incorporated into the corresponding
linear-programming formulations.

We consider both real and integer feasibility tests.
In the real LP relaxation, the enumerator coefficients
are treated as nonnegative real variables, and
integrality and divisibility constraints are omitted,
allowing more efficient computation.

\subsection{Quantum MDS Specialization}
\label{subsec:qmds_weight_distribution}

For an $[[n,k,d]]_q$ quantum MDS code with $k>0$, the stabilizer and its symplectic dual
are additive MDS codes over $\mathbb F_{q^2}$.
Their weight distributions are uniquely determined
by the code parameters~\cite{Tol94,Ald20,HG20}.
\begin{align}
	A_0&=1,\qquad A_j=0\quad \mbox{ for }1\le j<n-d+2,\nonumber\\
	A_j
	&=\binom{n}{j}(q^2-1)
	\sum_{\ell=0}^{j-n+d-2}
	(-1)^\ell\binom{j-1}{\ell}q^{2(j-n+d-2-\ell)},\nonumber\\
	 &\mbox{ for }j\ge n-d+2; \label{eq:MDS_A}\\
	B_0&=1,\qquad B_j=0\quad  \mbox{ for } 1\le j<d,\nonumber\\
	B_j
	&=\binom{n}{j}(q^2-1)
	\sum_{\ell=0}^{j-d}
	(-1)^\ell\binom{j-1}{\ell}q^{2(j-d-\ell)},\nonumber\\
	&\mbox{ for } j\ge d.\label{eq:MDS_B}
\end{align}

Moreover, the MDS property implies that the number
of codewords with a given support depends only on
the support size~\cite{Ald20,ElKhamyMcEliece05}.
Consequently, for any partition
$[n]=T\sqcup U$, with $|T|=t$ and $|U|=u$,
the split weight distributions are
uniquely determined by
\begin{align}
	A_{i,j}^{T,U}
	&=
	\frac{\binom{t}{i}\binom{u}{j}}
	{\binom{n}{i+j}}A_{i+j},	\label{eq:qmds_split_distribution}\\
	B_{i,j}^{T,U}
	&=
	\frac{\binom{t}{i}\binom{u}{j}}
	{\binom{n}{i+j}}B_{i+j}.
	\label{eq:qmds_split_distribution2}
\end{align}

\subsection{Computational Examples}
\label{subsec:LP_examples}

We illustrate the effectiveness of the refined LP
formulation through three levels of enumerator resolution:
\begin{align*}
	H \ \longrightarrow\ H\times R_X
	\ \longrightarrow\ R_X\times R_Y.
\end{align*}
For a reference stabilizer of weight $u$, the
$H\times R_X$ and $R_X\times R_Y$ enumerators
have $O((n-u+1)(u+1)^2)$ and
$O((n-u+1)^3(u+1)^3)$ coefficients, respectively.
The rapid growth in the number of variables
makes direct integer enumeration  computationally expensive.

We therefore adopt a two-stage computational approach.
First, we enumerate integer candidate weight distributions
at the Hamming level $H$, subject to the Shor--Laflamme
and shadow constraints, together with the three-type
counting constraints for binary stabilizer codes.
The surviving candidates are then tested against
higher-resolution constraints using real LP feasibility
checks, without requiring
integer enumeration.

For qudit stabilizer codes, we focus on quantum MDS codes,
whose weight distributions are uniquely determined by their
code parameters.  For general non-MDS codes, the
Hamming-level constraints often leave a prohibitively
large number of integer candidates.

\paragraph{Qubit stabilizer codes}

Table~\ref{tab:LP_survivors} summarizes the numbers
of candidate weight distributions surviving
the SLR3 constraints and subsequent refined LP tests.
Increasing the enumerator resolution generally
reduces the number of survivors.

For $[[13,5,4]]$ and $[[14,3,5]]$,
integer LP tests at the $H\times R_X$ level
give the same survivor counts as real LP tests
at the finer $R_X\times R_Y$ level.
Note that the nonexistence of $[[13,5,4]]$ was
previously established in~\cite{BFMP11}.

For $[[17,6,5]]$, partial $H\times R_X$ tests were completed for reference weights \(u=6,7,15,16,17\), reducing the SLR3  candidates from 292 to 267. A dash (--) indicates that the corresponding 
computations were attempted but not completed
due to their computational cost.

The resulting candidate sets can be further analyzed
using structural constraints, as discussed in
the next section.

\begin{table}[t]
	\centering
	\caption{Numbers of candidate weight distributions
		surviving the Shor--Laflamme, shadow, and three-type
		counting (SLR3) constraints and subsequent refined
		real LP feasibility tests. A dash (--) indicates
		that the computational search has not been completed.}
	\label{tab:LP_survivors}
	\begin{tabular}{cccc}
		\hline
		Code parameters & SLR3 & $H\times R_X$ & $R_X\times R_Y$ \\
		\hline
		$[[13,5,4]]$ & $30$   & $10$ & $9$ \\
		$[[14,3,5]]$ & $47$   & $4$  & $2$ \\
		$[[16,5,5]]$ & $95$   & $66$ & --  \\
		$[[17,3,6]]$ & $7$    & $7$  & --  \\
		$[[17,6,5]]$ & $292$  & $<268$   & --  \\
		$[[18,4,6]]$ & $2868$ & --  & --  \\
		$[[23,1,9]]$ & $1$    & --   & --  \\
		\hline
	\end{tabular}
\end{table}

\paragraph{Qutrit quantum MDS codes}

Next, we apply the quantum MDS specialization to
the parameter sets $[[11,5,4]]_3$ and $[[11,3,5]]_3$.
By \eqref{eq:MDS_A}, their stabilizer weight
enumerators are uniquely determined as
\begin{align*}
	[[11,5,4]]_3:\quad
	&A(x,y)=x^{11}+440x^2y^9+288y^{11};\\
	[[11,3,5]]_3:\quad
	&A(x,y)=x^{11}+1320x^3y^8+440x^2y^9\\
	&\hspace{1.5cm}
	+3168xy^{10}+1632y^{11}.
\end{align*}
However, both parameter sets can be excluded
using the full $H\times R_X$ refinement.

For $[[11,5,4]]_3$, we choose a weight-$9$
stabilizer element as a reference.
The corresponding full $H\times R_X$ LP,
subject to the Hamming projection constraints
\eqref{eq:qmds_split_distribution}
and \eqref{eq:qmds_split_distribution2},
is infeasible.

Similarly, for $[[11,3,5]]_3$, choosing a
weight-$8$ stabilizer reference also yields
an infeasible $H\times R_X$ LP.

Thus, the refined constraints exclude
both parameter sets, which survive the
Hamming-level LP. This improves two entries
in Grassl's code table~\cite{Grassl}.
We remark that their nonexistence was
recently established through the
classification of additive MDS codes
over $\mathbb F_9$~\cite{BGL23}.

\section{Structural Obstructions Beyond Linear Programming}
\label{sec:structural}

As in classical coding theory~\cite{MS77},
weight distributions satisfying LP constraints
need not correspond to realizable codes.
We therefore supplement the refined LP
constraints developed in the preceding section
with combinatorial arguments based on
shadow-coset counting and the simplex
mapping~\cite{CRSS98} to further exclude
candidate weight distributions.

\subsection{Pauli shadow-coset packing}

Let $\cS\subseteq\mathcal V_2^n$ be the
stabilizer subspace of an $[[n,k,d]]$
stabilizer code, and define
\begin{align*}
	\cS_0
	=\{\mbv\in\cS:\wt(\mbv)\equiv0\pmod2\}.
\end{align*}
If $\cS$ contains an odd-weight element,
then $\cS_0$ is a subspace of dimension
$n-k-1$.
The corresponding Rains shadow~\cite{Rains99b}
is the coset
\begin{align*}
	\operatorname{Sh}(\cS)
	=\cS_0^\perp\setminus\cS^\perp.
\end{align*}
Let $S_j$ denote the number of
weight-$j$ vectors in
$\operatorname{Sh}(\cS)$.

The SLR3 constraints leave $292$ integer
weight distributions  of Case~I  for $[[17,6,5]]$
in Table~\ref{tab:LP_survivors},
while Cases~II and~III are infeasible.
Since all $292$ candidates satisfy
$A_1=A_2=A_3=A_4=0$, every nonzero
element of $\cS^\perp$ for any
$[[17,6,5]]$ stabilizer code realizing
one of these distributions must have
weight at least five.
Moreover, since $\operatorname{Sh}(\cS)$
is a coset of $\cS^\perp$, the sum of
any two distinct shadow vectors belongs
to $\cS^\perp\setminus\{\mathbf0\}$.
Consequently, any two distinct shadow
vectors must have Pauli distance at
least five, imposing strong restrictions
on the number and arrangement of
low-weight shadow vectors.
This leads to the following packing bound.

\begin{lemma}[Low-weight shadow-coset packing]
	\label{lem:shadow_coset_distance_five}
	Suppose that an $[[n,k,5]]$
	stabilizer code exists with
	$A_1=A_2=A_3=A_4=0$.
Then one of the following two cases holds:
\begin{align*}
	\begin{cases}
		S_1=0,\quad S_3\le n;\\
		S_1=1,\quad S_3=0.
	\end{cases}
\end{align*}
\end{lemma}

\begin{proof}
	
	For any two distinct weight-$3$
	vectors $\mbu,\mbv\in
	\operatorname{Sh}(\cS)$, we have
	$\mbu+\mbv\in
	\cS^\perp\setminus\{\mathbf0\}$,
	so $\wt(\mbu+\mbv)\ge5$ since the distance is $5$ and $B_j=A_j=0$ for $1\leq j\leq 4$.
	
	Suppose their supports overlap in
	$r$ coordinates with identical
	nonidentity Pauli labels and
	$s$ coordinates with different
	nonidentity Pauli labels.
	Then
	\begin{align*}
		\wt(\mbu+\mbv)=6-2r-s\ge5.
	\end{align*}
	Hence $r=0$ and $s\leq 1$. This implies that
	two distinct weight-$3$ vectors
	in $\operatorname{Sh}(\cS)$
	cannot have the same nonidentity
	Pauli label at the same coordinate.
	
Since there are only three nonidentity
Pauli labels $X,Y,Z$, if a coordinate
appears in four or more weight-$3$
vectors, at least two of them must
have the same Pauli label at that
coordinate, a contradiction.
Thus each coordinate can appear in
at most three weight-$3$ vectors.
Counting their supports yields
$3S_3\le3n$, and hence $S_3\le n$.
	
	Similarly, two distinct weight-$1$
	vectors in $\operatorname{Sh}(\cS)$
	would have a sum of weight at most
	two, contradicting the minimum-weight
	condition on $\cS^\perp$.
	Thus $S_1\le1$.
	
	Finally, the sum of a weight-$1$
	vector and a weight-$3$ vector
	has weight at most four.
	Such vectors therefore cannot
	both belong to $\operatorname{Sh}(\cS)$,
	proving that $S_1>0$ implies $S_3=0$.
\end{proof}

\subsubsection{Nonexistence of $[[17,6,5]]$}
Suppose that a $[[17,6,5]]$
stabilizer code exists.
Its weight distribution must be
one of the $292$ candidates in Table~\ref{tab:LP_survivors}.

The shadow coefficients are determined
by the stabilizer weight enumerator through the shadow identity~\eqref{eq:qudit_shadow_hamming}
\begin{align*}
	S_j
	=\frac{1}{2^{11}}
	\sum_{w=0}^{17}(-1)^w
	A_w K_j^{(4)}(w;17),
\end{align*}
where $K_j^{(4)}$ is the
quaternary Krawtchouk polynomial~\eqref{eq:krawtchouk}.

Applying  Lemma~\ref{lem:shadow_coset_distance_five}
to the $292$ candidates gives
the exclusions summarized in
Table~\ref{tab:n17k6_exclusion_counts},
classified by the first violated condition.

\begin{table}[t]
	\centering
	\caption{Exclusion of the $292$ candidate weight
		distributions for $[[17,6,5]]$ stabilizer codes.}
	\label{tab:n17k6_exclusion_counts}
	\begin{tabular}{l|r}
		\hline
		Violated condition & Number of candidates\\
		\hline
		$S_3\le17$ & 258\\
		$S_1\le1$ & 1\\
		$S_1>0\Longrightarrow S_3=0$ & 33\\
		\hline
		Total & 292\\
		\hline
	\end{tabular}
\end{table}

Thus every candidate violates
Lemma~\ref{lem:shadow_coset_distance_five}.
Consequently, no $[[17,6,5]]$
binary stabilizer code exists.

\subsection{Simplex representation}
\label{sec:simplex_mapping}

We next introduce the simplex representation,
which maps binary Pauli-label codes to binary linear codes.

For $q=2$, let $\mbv=(\mbx,\mbz)\in\mathcal V_2^n$,
where $\mbx,\mbz\in\mathbb F_2^n$.
Consider the binary simplex map
$\xi:\mathcal V_2^n\longrightarrow\mathbb F_2^{3n}$
~\cite{CRSS98} defined by
\begin{align*}
	\xi(\mbx,\mbz)=(\mbx,\mbz,\mbx+\mbz).
\end{align*}
The action of the simplex map on the binary
Pauli labels is given by
\begin{equation}
	\begin{aligned}
		I&\leftrightarrow 00\mapsto000,\\
		X&\leftrightarrow 10\mapsto101,\\
		Z&\leftrightarrow 01\mapsto011,\\
		Y&\leftrightarrow 11\mapsto110.
	\end{aligned}
	\label{eq:simplex_symbols}
\end{equation}
Each nonidentity Pauli operator is mapped to a binary word
of weight two. Consequently,
\begin{align}
	\operatorname{wt}_H(\xi(\mbv))
	=2\wt(\mbv).
	\label{eq:simplex_weight}
\end{align}
Moreover,
\begin{align}
	\xi(\mbv)\cdot\xi(\mbv')
	=\langle\mbv,\mbv'\rangle.
	\label{eq:simplex_symplectic}
\end{align}
Thus, the commutation relation between two Pauli operators
is determined by the binary Euclidean inner product
of their simplex images.

Let $\cS\subseteq\mathcal V_2^n$ be the isotropic subspace
corresponding to an $[[n,k,d]]$ stabilizer code.
Then $C=\xi(\cS)$ is a binary self-orthogonal
$[3n,n-k]$ code.
Let $A_j=|\{\mbv\in\cS:\wt(\mbv)=j\}|$.
	By \eqref{eq:simplex_weight},  $C$ contains exactly $A_j$ codewords of weight $2j$
and no codewords of odd weight.
Thus, its Hamming weight enumerator is
\begin{align}
	W_C(x,y)=A(x^2,y^2).
	\label{eq:simplex_C_weight}
\end{align}
Moreover,	by the simplex inner-product identity
\eqref{eq:simplex_symplectic},
\begin{align}
	C^\perp=\xi(\cS^\perp)\oplus\mathrm{Ker},
	\qquad C^\perp/\mathrm{Ker}\simeq\cS^\perp,
	\label{eq:simplex_dual_quotient}
\end{align}
where 
\begin{align}
\mathrm{Ker}=\{(\mbr,\mbr,\mbr):\mbr\in\mathbb F_2^n\}. \label{eq:kernel}
\end{align}
One can verify that $\xi(\cS^\perp)\subseteq C^\perp$,
$\mathrm{Ker}\subseteq C^\perp$, and
$\xi(\cS^\perp)\cap \mathrm{Ker}=\{0\}$.

\begin{lemma}[Weight enumerator of the simplex dual]
	\label{lem:simplex_dual_weight}
	Let $\cS\subseteq\mathcal V_2^n$ be the isotropic subspace
	corresponding to an $[[n,k,d]]$ stabilizer code,
	and let $B(x,y)$ denote the Pauli weight enumerator
	of its symplectic dual $\cS^\perp$.
	Let $C=\xi(\cS)$.
	Then the Hamming weight enumerator
	of $C^\perp$ is given by
	\begin{align}
		W_{C^\perp}(x,y)
		=B(x^3+y^3,x^2y+xy^2).
		\label{eq:simplex_dual_weight}
	\end{align}
\end{lemma}

\begin{proof}

By 	\eqref{eq:simplex_dual_quotient},  $C^\perp$ is the disjoint union of the cosets
$\xi(\mbv)+\mathrm{Ker}$ over $\mbv\in\cS^\perp$.	
	By \eqref{eq:simplex_symbols} and \eqref{eq:kernel},
	each coordinate of a coset $\xi(\mbv)+\mathrm{Ker}$
	contains two possible binary symbols, obtained
	by adding either $000$ or $111$ to the
	corresponding simplex image.
	Their Hamming weights are $0$ and $3$
	for the identity Pauli symbol, and $1$ and $2$
	for each nonidentity Pauli symbol.

	Hence a Pauli vector of weight $j$ contributes
	\begin{align*}
		(x^3+y^3)^{n-j}(x^2y+xy^2)^j
	\end{align*}
	to the weight enumerator of $C^\perp$.
	Summing over all Pauli vectors in $\cS^\perp$
	gives \eqref{eq:simplex_dual_weight}.
\end{proof}

Applying the classical MacWilliams identity to
$C=\xi(\cS)$ and using \eqref{eq:simplex_dual_weight},
we obtain
\begin{align*}
	B(x^3+y^3,x^2y+xy^2)
	=\frac{1}{|\cS|}
	A((x+y)^2,(x-y)^2).
\end{align*}

\subsubsection{Nonexistence of EA stabilizer codes}
\label{subsec:EA_simplex_nonexistence}
We  establish a connection between
maximal-entanglement EA stabilizer codes~\cite{LBW13,LBW14}
and binary LCD codes~\cite{Massey92,DKO+17}
through the simplex map.
An $[[n,k,d;n-k]]$ maximal-entanglement EA code
encodes $k$ logical qubits into $n$ physical qubits
using $n-k$ shared ebits.
Recall that a binary linear code $C$ is LCD
if $C\cap C^\perp=\{\mathbf 0\}$.

\begin{lemma}
	The existence of an $[[n,k,d;n-k]]$
	maximal-entanglement EA stabilizer code
	implies the existence of a binary LCD
	$[3n,2k,2d]$ code.
\end{lemma}

\begin{proof}
	By \cite{LBW13,LBW14}, an $[[n,k,d;n-k]]$
	maximal-entanglement EA code is associated
	with an additive quaternary code
	$L\subseteq\mathcal V_2^n$ of size $2^{2k}$,
	minimum distance $d$, and
	$L\cap L^\perp=\{\mathbf 0\}$.
	By \eqref{eq:simplex_weight} and
	\eqref{eq:simplex_symplectic},
	$\xi(L)$ is a binary linear
	$[3n,2k,2d]$ code satisfying
	\begin{align*}
		\xi(L)\cap\xi(L)^\perp
		=\xi(L\cap L^\perp)
		=\{\mathbf 0\}.
	\end{align*}
	Hence $\xi(L)$ is an LCD code.
\end{proof}

The following weight-parity condition is the
$r=0$ case of the three-type counting identities
in~\cite[Theorem~16]{LA18}; see also
Lemma~\ref{lem:three_type_counting}
in Appendix~\ref{app:three_type_counting}.

\begin{corollary}[Weight-parity condition]
	\label{cor:EA_weight_parity}
	Let $L\subseteq\mathcal V_2^n$ be a binary
	linear subspace of dimension $2k$ satisfying
	$L\cap L^\perp=\{\mathbf0\}$.
	Then
	\begin{align*}
		\left|\sum_{\mbv\in L}(-1)^{\wt(\mbv)}\right|
		=2^k.
	\end{align*}
\end{corollary}

Under the binary simplex mapping,
$C=\xi(L)$ is a binary LCD code of dimension $2k$,
and the weight-parity condition becomes
$|W_C(1,i)|=2^k$.
This is equivalent to the Gauss-sum characterization
of binary LCD codes in~\cite{KX26},
and follows from the more general three-type
counting identities of~\cite{LA18}.

\paragraph{Nonexistence of $[[12,6,6;6]]$}
By the classification of additive $[12,6,6]_4$ codes
in~\cite[Theorem~2]{BierbrauerMP20},
the possible weight enumerators are
\begin{align*}
	W_{\mathrm I}(x,y)
	=&x^{12}+396x^6y^6+1485x^4y^8\\
	&+1980x^2y^{10}+234y^{12},\\
	W_{\mathrm {II}}(x,y)
	=&x^{12}+330x^6y^6+396x^5y^7
	+495x^4y^8 \notag\\
	& +1320x^3y^9+990x^2y^{10}
	+396xy^{11}+168y^{12}.
\end{align*}
Evaluating at $(x,y)=(1,-1)$ gives
\begin{align*}
	|W_{\mathrm I}(1,-1)|=4096,\qquad
	|W_{\mathrm {II}}(1,-1)|=128.
\end{align*}
neither of which satisfies the condition
$|W(1,-1)|=2^6=64$ required by
Corollary~\ref{cor:EA_weight_parity}.
Hence no $[[12,6,6;6]]$ EA stabilizer code exists.

\subsubsection{Nonexistence of $[[15,4,10;11]]$}
By the simplex mapping, an
$[[15,4,10;11]]$ EA stabilizer code would give
a binary LCD $[45,8,20]$ code $C=\xi(L)$.
Bouyuklieva and Bouyukliev classified all
binary linear $[45,8,20]$ codes into $424208$
equivalence classes, of which $424207$ are
doubly even~\cite[Example~1]{BouyuklievaBouyukliev20}.
Since every doubly-even binary linear code is
self-orthogonal, none of these codes can be LCD.

The remaining class has weight enumerator
\begin{align*}
	W_C(x,y)
	={}&x^{45}+99x^{25}y^{20}+90x^{23}y^{22}
	+15x^{21}y^{24} \notag\\
	&+45x^{17}y^{28}+6x^{15}y^{30}.
\end{align*}
Since $\operatorname{wt}_H(\xi(\mbv))=2\wt(\mbv)$,
the corresponding Pauli weight-parity sum is
\begin{align*}
	\left|\sum_{\mbv\in L}(-1)^{\wt(\mbv)}\right|
	&=\left|\sum_{\mbc\in C}
	(-1)^{\operatorname{wt}_H(\mbc)/2}\right|\\
	&=|1+99-90+15+45-6|=64.
\end{align*}
This violates Corollary~\ref{cor:EA_weight_parity},
which requires the absolute value to be $2^4=16$.
Therefore, no $[[15,4,10;11]]$ EA stabilizer code exists.

\subsubsection{Nonexistence of $[[17,13,4;4]]$}
By~\cite[Theorems~3--4]{BierbrauerMP20},
every additive $[17,13,4]_4$ code is equivalent
to the dual of the elliptic-quadric
$[17,4,12]_4$ code $E$, whose weight enumerator is
\begin{align}
	W_E(x,y)=x^{17}+204x^5y^{12}+51xy^{16}.
\end{align}
Applying the quaternary additive MacWilliams identity
to $L=E^\perp$ yields
\begin{align*}
	\sum_{\mbv\in L}(-1)^{\wt(\mbv)}
	&=W_L(1,-1) \notag
	=\frac{1}{|E|}W_E(-2,2)
	=-2^{17},
\end{align*}
while Corollary~\ref{cor:EA_weight_parity} requires an absolute value of $2^{13}$.
Thus no $[[17,13,4;4]]$ EA stabilizer code exists.

\subsubsection{Nonexistence of $[[19,3,14;16]]$}
Suppose that a $[[19,3,14;16]]$ EA
stabilizer code exists. By the simplex
mapping, its logical Pauli-label subspace
$L$ gives rise to a binary LCD
$[57,6,28]$ code $C=\xi(L)$.

However, Liu and Li showed that the
maximum minimum distance of binary LCD
codes of length $57$ and dimension $6$
is $27$~\cite[Table~2]{LiuLi24}.
Therefore, no $[[19,3,14;16]]$
EA stabilizer code exists.

\subsection{Binary shadow-coset counting}
\label{sec:shadow_coset_counting}

We now combine the simplex representation
with shadow-coset counting to derive
additional restrictions on stabilizer
weight distributions.

Let $C=\xi(\cS)\subseteq\mathbb F_2^{3n}$
be the simplex image of the stabilizer
subspace $\cS$, and let
$C_0=\xi(\cS_0)$, where $\cS_0$
is the even-weight subspace defined above.

The corresponding binary shadow coset is
\begin{align*}
	T=C_0^\perp\setminus C^\perp.
\end{align*}
If $\cS$ contains an odd-weight element,
then $C_0$ has codimension one in $C$.
Consequently,
$\dim C_0^\perp=\dim C^\perp+1$,
and $T$ is a single coset of $C^\perp$.

\subsubsection{Nonexistence of $[[14,3,5]]$}

In Table~\ref{tab:LP_survivors}, the refined
LP tests leave two candidate stabilizer
weight distributions, labeled $17$ and $35$:
\begin{align*}
	A^{(17)}(x,y)
	={}&x^{14}+x^{10}y^4+5x^8y^6
	+32x^7y^7 \notag\\
	&+199x^6y^8+400x^5y^9
	+350x^4y^{10} \notag\\
	&+384x^3y^{11}+463x^2y^{12}
	+208xy^{13}+5y^{14},\\
	A^{(35)}(x,y)
	={}&x^{14}+x^{10}y^4+4x^8y^6
	+34x^7y^7 \notag\\
	&+201x^6y^8+394x^5y^9
	+350x^4y^{10} \notag\\
	&+390x^3y^{11}+461x^2y^{12}
	+206xy^{13}+6y^{14}.
\end{align*}
Both candidates satisfy $A_4=1$;
that is, each has exactly one
weight-4 stabilizer element.

For either candidate, the simplex image
$C=\xi(\cS)$ would be a
self-orthogonal $[42,11]$ binary code.
By Lemma~\ref{lem:simplex_dual_weight}
and the quantum MacWilliams identity
\eqref{eq:qudit_hamming_MI},
both candidates yield
\begin{align*}
	W_{C^\perp}(x,y)
	&=B(x^3+y^3,x^2y+xy^2) \notag\\
	&=x^{42}+14x^{39}y^3
	+x^{38}y^4+\cdots.
\end{align*}
Hence, in either case, $C^\perp$ contains
no weight-2 codewords, exactly $14$
weight-3 codewords, and exactly one
weight-4 codeword.

We next determine the number of
weight-$3$ words in the binary shadow coset $T$.
For both candidates, $|C_0|=2^{10}$,
and
\begin{align*}
	W_{C_0}(x,y)
	=\sum_{\substack{w=0\\w\text{ even}}}^{14}
	A_w x^{42-2w}y^{2w}.
\end{align*}
By the classical MacWilliams identity,
the number of weight-3 codewords
in $C_0^\perp$ is
\begin{align*}
\frac{1}{2^{10}}
	\sum_{\substack{w=0\\w\text{ even}}}^{14}
	A_w K_3^{(2)}(2w;42),
\end{align*}
where $K_3^{(2)}(t;42)$ is the binary
Krawtchouk polynomial  \eqref{eq:krawtchouk}  of degree three.
Substituting the two candidate
distributions gives $34$ and $33$
weight-3 codewords in $C_0^\perp$,
respectively.
Since $T=C_0^\perp\setminus C^\perp$
and $C^\perp$ contains exactly $14$
weight-$3$ codewords, the two
candidates require $T$ to contain
$20$ and $19$ weight-$3$ words,
respectively.

We show that these counts are impossible
under the low-weight structure of
$C^\perp$.

\begin{lemma}
	\label{lem:triple_packing_42}
	Suppose that a $[[14,3,5]]$ stabilizer
	code exists, and let
	$C\subseteq\mathbb F_2^{42}$ be
	its simplex image, with
	$T=C_0^\perp\setminus C^\perp$
	the corresponding shadow coset.
	If $C^\perp$ contains no weight-$2$
	codewords and exactly one weight-$4$
	codeword, then $T$ contains at most
	$14$ weight-$3$ words.
\end{lemma}

\begin{proof}
	Let $\cX$ denote the set of supports
	of weight-$3$ words in $T$.
	For any two distinct weight-$3$ words
	$\mbu,\mbv\in T$, we have
	$\mbu+\mbv\in C^\perp\setminus\{\mathbf0\}$.
	Since $C^\perp$ contains no weight-$2$
	codewords, their supports cannot
	overlap in two coordinates.
	
	If all weight-$3$ words in $T$ are
	pairwise disjoint, then
	$|\cX|\le\lfloor42/3\rfloor=14$.
	
	Otherwise, let $\mbu,\mbv\in T$
	be two weight-$3$ words whose supports
	overlap in exactly one coordinate.
	Then $\mbu+\mbv$ is the unique
	weight-$4$ codeword in $C^\perp$.
	
	Consider any other weight-$3$ word
	$\mbt\in T\setminus\{\mbu,\mbv\}$.
	If $\mbt$ overlaps $\mbu$ in one
	coordinate, then
	$\mbu+\mbt=\mbu+\mbv$,
	implying $\mbt=\mbv$, a contradiction.
	The same argument applies to $\mbv$.
	Thus $\mbt$ must be disjoint from
	both $\mbu$ and~$\mbv$.
	
	Furthermore, any two such remaining
	weight-$3$ words $\mbt,\mbs$ must
	be disjoint. Otherwise, their sum
	$\mbt+\mbs$ would be the unique
	weight-$4$ codeword $\mbu+\mbv$,
	whose support is contained in
	$\operatorname{supp}(\mbu)\cup
	\operatorname{supp}(\mbv)$.
	This contradicts the fact that
	$\mbt$ and $\mbs$ are disjoint
	from both $\mbu$ and $\mbv$.
	
	Since
	$|\operatorname{supp}(\mbu)\cup
	\operatorname{supp}(\mbv)|=5$,
	the remaining weight-$3$ words
	have pairwise disjoint supports
	contained in the other $37$
	coordinates. Therefore,
	\begin{align*}
		|\cX|
		\le 2+\left\lfloor\frac{42-5}{3}\right\rfloor
		=14.
	\end{align*}
\end{proof}

The required counts of $20$ and $19$
weight-$3$ words in $T$ both violate
Lemma~\ref{lem:triple_packing_42}.
Thus neither candidate is realizable,
and no $[[14,3,5]]$ binary stabilizer 
code exists.

\subsection{Remaining candidate sets}

For $[[16,5,5]]$, the initial SLR3
enumeration gives $83$ candidates in
Case~I and $12$ in Case~II, while
Case~III is infeasible.
The refined LP and structural tests
reduce these to $18$ and $9$ candidates,
respectively.
Thus $27$ candidates remain, and
nonexistence has not been established.

For $[[18,4,6]]$, the initial SLR3
enumeration gives $2835$ candidates
in Case~I and $33$ in Case~II.
Shadow-coset pair counting reduces
Case~I to $882$ candidates, while
all $33$ candidates in Case~II remain.
Hence $915$ candidates survive.

For $[[17,3,6]]$, seven candidate
weight distributions remain, and
the available structural constraints
have not established nonexistence.

For $[[23,1,9]]$, only one candidate
weight distribution remains.
This distribution satisfies
$A_1=\cdots=A_8=0$ and $A_w=0$
for all odd $w$.
Thus any realizing stabilizer code
would be nondegenerate and have
$\cS_0=\cS$, so its Rains shadow
coset is empty.
Consequently, the shadow-coset
packing argument developed above
does not apply directly.
Further structural constraints
are needed to determine whether
this distribution is realizable.

\section{Conclusion}
\label{sec:conclusion}

We developed a unified Fourier formulation of
quantum weight enumerators and their
MacWilliams and shadow identities.
Within this framework, we introduced
twisted Fourier transforms extending the
ordinary MacWilliams and shadow identities.
The two-class reduction,
Theorem~\ref{thm:two_class_reduction},
shows that all $2q^2$ twisted Fourier
transforms are Weyl translations of either the MacWilliams or shadow transform.
However, projection compatibility shows
that compatible twists yield no independent
coefficientwise positivity constraints
within a common refinement.

Despite this limitation, our unified
framework naturally yields refined
MacWilliams~\cite{CRSS98,LA18} and shadow
identities with stronger LP constraints.
We demonstrate their effectiveness through
two qutrit quantum MDS examples.
The simplex representation provides
additional structural constraints for
four maximally entangled binary EA codes.
Furthermore, shadow-coset counting,
both directly in Pauli space and through
the simplex representation, establishes
the nonexistence of binary stabilizer
codes with parameters $[[17,6,5]]$
and $[[14,3,5]]$.

For binary quantum codes, we show that
the $010$ twisted autocorrelation
admits an operational interpretation
through Bell sampling on two identical
copies of a quantum state.
The translation relation between the
$010$ and $111$ transforms further connects
Bell-sampling statistics to the
Rains shadow enumerator.

We conclude by discussing several
limitations of our current approach
and directions for future research.

\paragraph{Scalability}
Our computational approach relies on
a manageable number of integer LP survivors.
For larger binary parameters, exhaustive
SLR3 enumeration often becomes prohibitive,
limiting subsequent refined LP and
structural tests.

For nonbinary stabilizer codes,
the number of LP survivors is
typically even larger.
We therefore focused on selected quantum
MDS parameters, including
$[[11,5,4]]_3$ and $[[11,3,5]]_3$,
as well as several cases over
$q=4,5,7$.
However, our techniques yielded no
additional nonexistence results
for these nonbinary cases.

\paragraph{Beyond character twists}
\label{subsec:positive_map_open}

Our two-class reduction shows that
compatible twisted Fourier transforms
yield no independent coefficientwise
positivity constraints within the same
refinement.
However, this result does not rule out
stronger constraints from twisted
transforms that require additional
enumerator variables.

\paragraph{Positive maps}
Another direction is to consider
positive maps.
The generalized Shor--Laflamme
enumerators~\cite{SL97} allow two
operators, which may be chosen as
$P$ and $\mathcal E(P)$, where
$\mathcal E$ is a positive map.

For the depolarizing family
\begin{align*}
	\mathcal D_\epsilon(A)
	=\epsilon A+(1-\epsilon)\frac{\Tr(A)}q I
\end{align*}
with $-\frac{1}{q-1}\le \epsilon\le1,$
we obtain an analytical redundancy
result from the decomposition
\begin{align*}
	\mathcal D_\epsilon
	=\alpha_\epsilon \mathcal D_1
	+(1-\alpha_\epsilon )\mathcal D_{-1/(q-1)},
\end{align*}
where
$\alpha_\epsilon =\frac{1+(q-1)\epsilon}{q}$
and $0\le\alpha_\epsilon\le1$.
The endpoints are the identity and
the normalized universal inversion.
By convexity, intermediate maps yield
no additional positivity constraints
beyond those of the endpoints.
Note that this argument does not directly apply
to arbitrary tensor products of
local depolarizing maps.

More general positive maps introduce
additional difficulties.
Maps not diagonal in the Weyl basis
generally produce cross terms
$\lambda(\mbv)\lambda(\mbw)^*$
with $\mbv\ne\mbw$, which are not
determined by the Weyl power-spectrum coefficients
$\Lambda(\mbv)$ and may require
additional variables.

We also explored amplitude-damping
and rotated-dephasing channels,
Breuer--Hall maps~\cite{Breuer06,Hall06},
and positive projections onto spin
factors~\cite{Stormer16}.
These investigations yielded no
certified improvement over existing
refined LP constraints, but no general
redundancy theorem was established.
Whether more general positive maps
can yield independent inequalities
involving only existing refined
enumerators remains open.

\paragraph{Nonbinary counting constraints}
The three-type counting constraints
of \cite[Theorem 16]{LA18} apply
to binary weight enumerators without
additional variables.
Related arithmetic constraints have
also been developed for ternary LCD
codes~\cite{KX26}.
Although quadratic-character counting
identities extend to nonbinary Weyl
groups, their counts are generally
not determined by ordinary Hamming
weight enumerators.
Whether analogous constraints involving
only the Shor--Laflamme coefficients
$A_j$ exist for nonbinary stabilizer
codes remains open.

\paragraph{Shadow-coset counting}
Our nonexistence results show that
shadow-coset packing and simplex-based
counting can exclude weight distributions
satisfying refined LP constraints.
A natural question is whether these
techniques can yield systematic
realizability constraints beyond
MacWilliams and shadow positivity.

In particular, the Pauli shadow
of a nondegenerate stabilizer code
inherits a minimum-distance restriction
from the dual code.
This suggests more general packing
bounds involving multiple low weights
and their pairwise intersections,
possibly combined with simplex
representations.
Whether such constraints can be
incorporated directly into the LP
framework remains open.

\paragraph{Signed weight enumerators}
Another possible extension concerns signed
weight enumerators used in magic state
distillation~\cite{rall2017signed,kalra2025invariant}.
Unlike the Weyl power spectrum, these
enumerators may retain phase information
that is lost upon taking squared magnitudes.
Developing a Fourier framework for such
phase-sensitive enumerators is an interesting question.

\section*{AI Disclosure}

OpenAI ChatGPT and Anthropic Claude were used
to assist with mathematical exploration,
computational code development and verification,
and manuscript preparation.
In particular, ChatGPT helped identify
a simple counting argument that excludes
all 292 LP candidates for $[[17,6,5]]$,
avoiding the formidable task of
examining them individually.
The authors take full responsibility for
the results and content of this work.

\bibliographystyle{IEEEtran}
\bibliography{IEEEabrv,../qecc}

\appendices

\section{An Alternative Proof of the Three-Type Counting Identities}
\label{app:three_type_counting}

\begin{lemma}[Three-type counting {\cite[Theorem~16]{LA18}}]
	\label{lem:three_type_counting}
	Let $V\subseteq\mathcal V_2^n$ be a binary
	linear subspace, and let $R=V\cap V^\perp$.
	Suppose $\dim V=2c+r$ and $\dim R=r$.
	Let $N_{\mathrm{even}}(V)$ and
	$N_{\mathrm{odd}}(V)$ denote the numbers of
	even- and odd-weight vectors in $V$, respectively.
	Then one of the following three cases holds:
	\begin{align*}
		\text{I:}\quad&
		N_{\mathrm{even}}(V)=2^{r-1}(4^c+2^c),\\
		&
		N_{\mathrm{odd}}(V)=2^{r-1}(4^c-2^c);
		\nonumber\\
		\text{II:}\quad&
		N_{\mathrm{even}}(V)=N_{\mathrm{odd}}(V)
		=2^{2c+r-1};
		\nonumber\\
		\text{III:}\quad&
		N_{\mathrm{even}}(V)=2^{r-1}(4^c-2^c),\\
		&
		N_{\mathrm{odd}}(V)=2^{r-1}(4^c+2^c).
	\end{align*}
	Case II is possible only when $r>0$.
\end{lemma}

\begin{proof}
	For any $\mbu,\mbv\in\mathcal V_2^n$,
	the Pauli weight satisfies~\cite{Rains99b}
	\begin{align*}
		\wt(\mbu)+\wt(\mbv)
		=\wt(\mbu+\mbv)+\langle\mbu,\mbv\rangle
		\pmod 2.
	\end{align*}
	Hence,
	\begin{align*}
		\left(
		\sum_{\mbv\in V}(-1)^{\wt(\mbv)}
		\right)^2
		&=
		\sum_{\mbu,\mbv\in V}
		(-1)^{\wt(\mbu)+\wt(\mbv)}\\
		&=
		\sum_{\mbu,\mbv\in V}
		(-1)^{\wt(\mbu+\mbv)+\langle\mbu,\mbv\rangle}\\
		&\stackrel{(a)}{=}
		\sum_{\mbu,\mbt\in V}
		(-1)^{\wt(\mbt)+\langle\mbu,\mbt\rangle}\\
		&=
		\sum_{\mbt\in V}(-1)^{\wt(\mbt)}
		\sum_{\mbu\in V}
		(-1)^{\langle\mbu,\mbt\rangle}\\
		&\stackrel{(b)}{=}
		|V|\sum_{\mbt\in R}(-1)^{\wt(\mbt)},
	\end{align*}
	where $(a)$ follows by setting
	$\mbt=\mbu+\mbv$ and
	$\langle\mbu,\mbu\rangle=0$
	for any $\mbu\in\mathcal V_2^n$;
	$(b)$ follows from character orthogonality
	on the subspace $V$,
	\begin{align*}
		\sum_{\mbu\in V}
		(-1)^{\langle\mbu,\mbt\rangle}
		=|V|\mathds{1}_{\{\mbt\in V^\perp\}}.
	\end{align*}

	If $R$ contains an odd-weight vector, then
	\begin{align*}
		\sum_{\mbt\in R}(-1)^{\wt(\mbt)}=0.
	\end{align*}
	Consequently,
	$		N_{\mathrm{even}}(V)
	=N_{\mathrm{odd}}(V)=2^{2c+r-1},$ which proves Case II. 
	Otherwise, every vector in $R$ has even
	weight, and
	\begin{align*}
		\left(
		\sum_{\mbv\in V}(-1)^{\wt(\mbv)}
		\right)^2
		=|V||R|=2^{2c+2r}.
	\end{align*}
	Thus
	$N_{\mathrm{even}}(V)-N_{\mathrm{odd}}(V)
	=\pm2^{c+r}.$
	Combining this with
	$N_{\mathrm{even}}(V)+N_{\mathrm{odd}}(V)
	=2^{2c+r}$ gives
	\begin{align*}
		N_{\mathrm{even}}(V)
		&=2^{r-1}(4^c\pm2^c),\\
		N_{\mathrm{odd}}(V)
		&=2^{r-1}(4^c\mp2^c),
	\end{align*}
	corresponding to Cases I and III.
	
	When $r=0$, $R=\{\mathbf0\}$,
	so Case II cannot occur.
\end{proof}

\end{document}